\newcommand{\be}{\begin{equation}}
\newcommand{\ee}{\end{equation}}
\newcommand{\bcl}{\begin{lstlisting}}
\newcommand{\ecl}{\end{lstlisting}}
\newcommand{\dd}{{\rm d}}

\documentclass[useAMS,usenatbib]{mn2e}
\usepackage{graphicx}
\usepackage{listings}
\newlength{\myfix}
\newlength{\myofix}
\def\ltsima{$\; \buildrel < \over \sim \;$}
\def\simlt{\lower.5ex\hbox{\ltsima}}   
\def\gtsima{$\; \buildrel > \over \sim \;$}
\def\gtsim{\lower.5ex\hbox{\gtsima}}

\title[Relative Pressure SPH] {{\em rpSPH}: a novel Smoothed Particle Hydrodynamics Algorithm} \author[Tom Abel]{Tom Abel\thanks{E-mail: tabel@stanford.edu} \\
  Kavli Institute for Particle-Astrophysics and Cosmology,
  Stanford University, SLAC National Accelerator Laboratory, Menlo Park 94025, USA}
\begin{document}

\date{to be submitted}
\pagerange{\pageref{firstpage}--\pageref{lastpage}} \pubyear{2010}
\maketitle
\topmargin=-1.6cm              % to fit on a US letter page
\label{firstpage}

\begin{abstract}
  We suggest a novel discretisation of the momentum equation for
  Smoothed Particle Hydrodynamics (SPH) and show that it significantly
  improves the accuracy of the obtained solutions. Our new formulation
  which we refer to as relative pressure SPH, {\em rpSPH}, evaluates
  the pressure force in respect to the local pressure. It respects
  Newtons first law of motion and applies forces to particles only
  when there is a net force acting upon them. This is in contrast to
  standard SPH which explicitly uses Newtons third law of motion
  continuously applying equal but opposite forces between particles.
  {\em rpSPH} does {\em not} show the unphysical particle noise, the
  clumping or banding instability, unphysical surface tension, and
  unphysical scattering of different mass particles found for standard
  SPH.  At the same time it uses fewer computational operations.
  and only changes a single line in existing SPH codes.
  We demonstrate its performance on isobaric uniform density
  distributions, uniform density shearing flows, the Kelvin--Helmholtz
  and Rayleigh--Taylor instabilities, the Sod shock tube, the
  Sedov--Taylor blast wave and a cosmological integration of the Santa
  Barbara galaxy cluster formation test. {\em rpSPH} is an improvement
  these cases. The improvements come at the cost of giving up exact
  momentum conservation of the scheme. Consequently one can also
  obtain unphysical solutions particularly at low resolutions.
%  We furthermore discuss how this formulation allows to study viscous
%  flows, is robust even with widely varying particles masses and
%  successfully apply the same principles to discretising the magnetic
%  forces in the ideal MHD limit.
\end{abstract}


\section{Motivation}

The smoothed particle hydrodynamics method was invented by
\cite{1977AJ.....82.1013L} and \cite{1977MNRAS.181..375G}, both with
interests in astrophysical applications. Besides an enormous
literature of successful application also many shortcomings of it have
been presented in the literature \citep{1993A&A...268..391S,
  1994MmSAI..65.1013H, Swegle1995123, 2002ApJ...569..501I,
  2007MNRAS.380..963A, 2009arXiv0906.0774R} often suggesting a fix to
the reported problem \citep[][to name but a few]{ 
Monaghan1994399,
Cummins1999584,
2000PThPS.138..609R,
2007A&A...464..447A,
Hu2007264,
2008IJNMF..56.1261G,
2008JCoPh.22710040P,
2009ApJ...701.1269B,
Rafiee20092785,
Xu20096703}
Similarly there have been troubling news of how seemingly small
differences in the initial setup led to very unexpected results
\citep[e.g.][]{Lombardi1999687}. 
This is all somewhat surprising given that in many of the cases where
large inaccuracies have been found the only relevant equation (besides
moving the particles $\dot{\vec{r}}=\vec{v}$) stems from the pressure
gradient accelerations
\be
\rho \frac{D\vec{v}}{Dt} = - \nabla p,
\ee
where $D/Dt$ denotes the Lagrangian derivative, and $p$ the pressure.  
In what follows we describe a new discretization of the momentum
equation that avoids essentially all of the previously known
problems of SPH. We will refer to this new method as ``relative pressure SPH''
or abbreviated as {\em rpSPH}. We will first describe it, discuss
implementation details and then present results for relevant tests
highlighting the superior performance of this new approach.

All the simulations shown here are carried out with
Gadget-2~\citep{2005MNRAS.364.1105S} (version 2.0.4) with only most
minor changes explained in the text.  The appendix describes how to
convert Gadget to our {\em rpSPH} formalism.

\section{Relative Pressure SPH}

The equation of motion without viscous or gravitational forces in
essentially all SPH codes and Gadget-2 \citep{2002MNRAS.333..649S,
  2005MNRAS.364.1105S}, in particular, is \be \frac{\dd \vec{v}_i}{\dd
  t} = - \sum_{j=1}^N m_j \left[ f_i \frac{P_i}{\rho_i^2} \nabla_i
  W_{ij}(h_i) + f_j \frac{P_j}{\rho_j^2} \nabla_i W_{ij}(h_j) \right],
\label{eqnmot} 
\ee where the $f_i$ are defined by \be f_i = \left[ 1 +
  \frac{h_i}{3\rho_i}\frac{\partial \rho_i}{\partial h_i} \right]^{-1}
\, , \ee and the abbreviation $W_{ij}(h)= W(|\vec{r}_{i}-\vec{r}_{j}|,
h)$ has been used for the kernel function, $W$. It employs variable
smoothing lengths so that the number of neighbours for each particle
with $|\vec{r}_{ij}| \le h_i$ is maintained at a nearly fixed value
$N_{\rm sph}$. The compact cubic spline kernel is used which is
summarized \cite{1992ARA&A..30..543M}, extends to radii as large
as the smoothing length $h$ and is zero outside. While many choices
would exist to use a different discretisation here
\citep{1992ARA&A..30..543M, 2009NewAR..53...78R} most previous work we
have found essentially retains a form very close to
equation~(\ref{eqnmot}) or uses another symmetric form that sums over
$(P_i+P_j)(\rho_i \rho_j)$. The reason previously given for these
choices are their symmetric form encapsulating Newtons third law of
motion, the action-reaction law. By guaranteeing that particles give
pairwise identical but reversed forces one ensures linear momentum
conservation of the entire scheme. The key here is that particles are
always pushing as soon as they have any pressure regardless of whether
there is a pressure gradient. The ones with the highest pressure
values are pushing the most. When one has a large number of particles
in a perfectly symmetric configuration all the pushing will average
out for an individual particle. This is to some extent what happens in
real gases. The pressure itself is mediated by the collision of the
molecules the gas is made of.

From a physical point of view of a Lagrangian fluid element, however,
one should only be interested in the pressure forces of neighbouring
fluid elements exerted on oneself, since the actual equation of motion
is $\rho\dot{v}=\nabla P$. This is the main idea of {\em rpSPH}, a
particle is accelerated only if a force is acting upon it,
i.e. Newtons first law of motion.  {\em rpSPH} derives its equation of
motion directly from equation~(\ref{eqnmot}) by subtracting the
constant pressure of the particle under consideration from the
pressures of all the particles being summed over. Since a gradient is
computed, the subtraction of a constant does not change the
mathematical meaning of the difference equation. However, as we will
demonstrate it dramatically affects the error properties of the entire
scheme.  The resulting equation of motion reads \be \frac{\dd
  \vec{v}_i}{\dd t} = - \sum_{j=1}^N m_j \left[
  \frac{P_j-P_i}{\rho_j^2} {\nabla_i W_{ij}(h_i)} \right].
\label{eqnmotrpSPH} 
\ee  One immediately notices that this
formulation breaks the symmetry between the pairwise forces of
particles. Particles that have a pressure difference are both
accelerated into the same direction along the pressure
gradient. Linear momentum conservation hence will only be achieved if
the modeled pressure gradients are resolved. On the other hand if one
does not resolved the relevant pressure gradients one cannot possibly
get a correct solution to a hydrodynamic problem in any case.

After all, it is important to recall that when constructing
conservative schemes one does not necessarily minimize the numerical
errors but rather ensures that one is making symmetric errors so that
the conserved quantity does not change. Consequently, in {\em rpSPH}
monitoring the total angular and linear momentum is an indicator of
whether one may have resolved the relevant length scales.

Many of the advantages of the entropic function based SPH formalism
\cite{2002MNRAS.333..649S} stem from avoiding the $P\,dV$ term that
generally is discretized analogous to equation~(\ref{eqnmot}). So in
this formalism {\em rpSPH} is particularly trivial to implement. It
involves setting the first term on the right hand side of
equation~(\ref{eqnmot}) to zero and change the second by subtracting
the pressure of the particle under consideration. This literally is
achieved by modifying one line of code in Gadget-2
\cite{2005MNRAS.364.1105S} as shown in the last appendix.  The
resulting scheme saves two multiplies, one division and one addition
for one additional subtraction in the main loop over neighbors. So
there is no performance penalty in using {\em rpSPH} as compared to
standard SPH.

{\em rpSPH} is seemingly close to equation~(3.1) of
\cite{1992ARA&A..30..543M} first discussed by
\cite{1996PASA...13...97M} which we will refer to as the Morris
formulation. It reads,
\be \frac{\dd
  \vec{v}_i}{\dd t} = - \sum_{j=1}^N m_j \left[
  \frac{P_j-P_i}{\rho_i \rho_j } {\nabla_i W_{ij}(h_i)} \right]. \ee

Monaghan dismissed his version for two reasons. The first is that ``an
isolated pair of particles with different pressures would bootstrap
themselves to infinity'' and the second is that it is difficult to
construct a consistent energy equation.  The latter is irrelevant in
the formalism evolving an entropic function
\cite{1977AJ.....82.1013L, 2002MNRAS.333..649S} in which the $PdV$\,
work does not enter.  The first reason we find unappealing since it is
actually the correct solution. The simulation having two particles
estimates a pressure gradient. So over the model volume, i.e. the two
particles and their smoothing volumes there exists a monotonic
pressure gradient. Both particles hence should be accelerated along
it. Interestingly, Monaghan did not discuss the equivalent case for
the symmetric standard SPH. In this case both particles push each
other to infinity no matter what. If they have the identical initial
pressures their center of mass will not change if they vary their
center of mass moves exactly as in {\em rpSPH}. In {\em rpSPH} they
will move together while in SPH they will accelerate each other apart
to infinity.  We have tested this on a spherical blob of material in
vacuum. We set the pressure of the particles after the densities have
been computed from kernel smoothing. This way all particles have
identical initial pressure. The configuration is completely stable in
{\em rpSPH} yet blows itself apart in SPH in just a sound crossing
time. The reason why we do not choose equation~(3.1) of
\cite{1992ARA&A..30..543M} is because we find it to be unstable at
least with the leapfrog time integrator in Gadget-2 (see
Figure~\ref{fig:mSPH-RT} below). Another formulation close to {\em
  rpSPH} discretization we could find in the literature is presented
\cite{262347}, who chose to subtract a background pressure. This still
leaves an equation of motion in which the pressure of the particle
under consideration remains part of the hydro force estimate.

An easy way to see why our discretization is valid (Wadsley, 2010,
private communication) recognizes $m_j/\rho_j\equiv dV$ as the volume
element $dV$ and sees that equation~\ref{eqnmotrpSPH} is equivalent to
$\int \left[ \nabla P\rho^{-1} - P \nabla \rho^{-1}\right] dV$, which
is the same as $\int \rho^{-1} \nabla P$ and is the term we want. The
previous version discussed by \cite{1996PASA...13...97M} in contrast
is the discretized form $\rho^{-1}\nabla P$ rather than our $\nabla
P\rho^{-1} - P \nabla \rho^{-1}$. Note that our form is also not the
general form suggested by equation~$2.13$ of
\cite{2005RPPh...68.1703M} and in this regard is a new formulation.
The striking aspect is that in the actual difference form all that is
new is that one index that used to be $i$ is now $j$. So this
literally is a one letter change to codes that implement the Morris
formulation. How this can lead to a dramatic change in accuracy
becomes  obvious from the standard error analysis.
\cite{PriceThesis2004} gives the error of the summation interpolant in
equation~(3.9) and the error of the gradient operator in
equation~(3.11). In a formulation as by \cite{1996PASA...13...97M} one
discretises $\rho_i^{-1} \nabla P$ and sees that the errors in the
interpolation of $\rho^{-1}$ and the errors of the chosen $\nabla P$
discretisation multiply. What \cite{1996PASA...13...97M} realized was
that there are no error terms in his discretisation for constant
functions in the gradient operator. However, the complete error terms
still end up being the product of the density estimate and the
pressure gradient estimate.  The advantage of the {\em rpSPH}
discretisation is that its error terms are the one of a gradient and
are {\em not} further multiplied by errors of a density
interpolation. It also retains the vanishing error for constant
functions of the Morris discretisation. 

Interestingly, a linear stability analysis reveals that {\em rpSPH}
has the same dispersion relation as the form of
\cite[][his equation~10]{1996PASA...13...97M}. He has shown that this form 
\begin{itemize}
\item is always stable, independent of the background pressure, 
\item has a numerical sound speed that depends less on the particle spacing
  as compared to standard SPH,
\item and does not have unphysical unstable transverse waves in two nor
  three dimensions when using kernels with compact support.
\end{itemize}
{\em rpSPH} retains all of these advantages while at the same time
reducing the discretisation error. 

In the standard SPH approach there are in fact infinitely many
possible choices for the discretisation of the pressure equation
\citep[equation 3.5 in][]{1992ARA&A..30..543M}. This is also true for
{\em rpSPH}. E.g. $P^{\sigma-1} \nabla P^\sigma/\rho - P \nabla 1/\rho =
\sigma \rho^{-1} \nabla P$ which suggest the discretisation
\be \frac{\dd
  \vec{v}_i}{\dd t} = - \sum_{j=1}^N m_j \sigma^{-1} \left[
  \frac{P_j^\sigma-P_i^\sigma}{P_i^{\sigma-1}\rho_j^2} {\nabla_i
    W_{ij}(h_i)} \right] \ee for any $\sigma$ different from zero. We have
verified that many choices of $\sigma$ work for a variety of test
problems. In the following, however, we restrict our attention to the
case of $\sigma=1$.

Whether these theoretically advantageous properties of {\em rpSPH}
hold up in practice is assessed in a range of test problems in the
following section. 

\section{Tests of rpSPH}

We will employ a Courant factor of $0.3$ (i.e. $0.15$ in Gadget where
the kernel has a maximal radius of $h$). 

\subsection{Reduced Velocity Noise}

\begin{figure}
\includegraphics[width=0.47\textwidth]{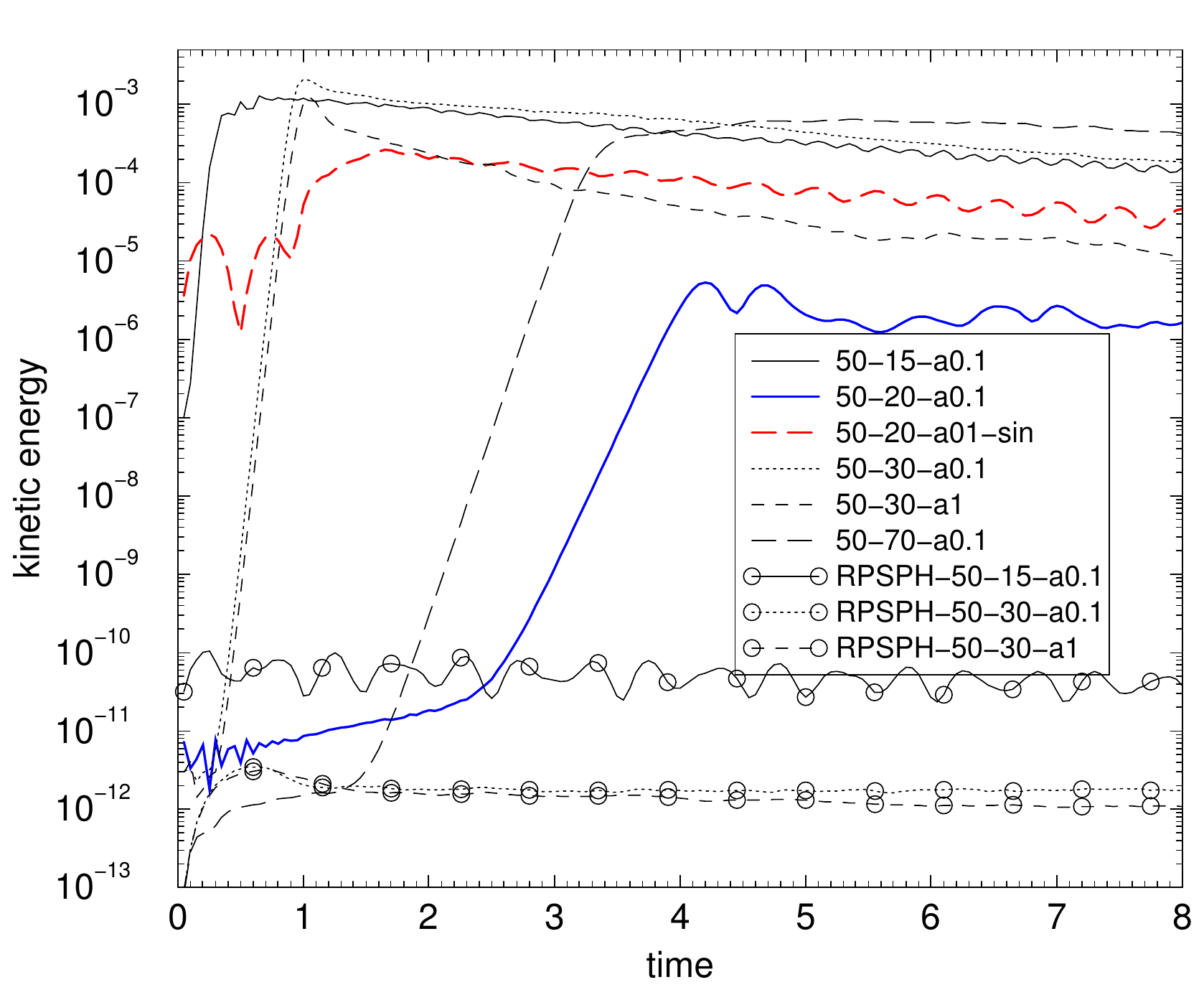}
\caption{The total kinetic energy in the static uniform density test
  as a function of time. The labels give the square root of the number
  of particles used, the number of neighbors and the artificial
  viscosity parameter. So the 50-30-a0.1 used $50^2$ particles, with
  30 neighbors and and artificial viscosity parameter $\alpha=0.1$. We
  clearly see that using more neighbors leads to a slower growth of
  the unphysical kinetic energy. Using higher viscosity values can not
  stop that growth but does influence how quickly this noise is damped
  over time. {\em rpSPH} even in the lowest numbers of neighbors
  reduces the kinetic energies associated with these unphysical
  motions by at least seven orders of magnitude and for sufficient
  neighbor numbers keeps it at zero to machine precision. }\label{fig:ke-noise}
\end{figure}

We start with $50^2$ particles on a periodic regular lattice with
$\gamma=1.4$, a sound speed and uniform density of unity and zero
initial velocities. The particles should stay at rest. However, as we
can see in Figure~\ref{fig:ke-noise} the total kinetic energy in the
volume grows rapidly. The total energy in the system is, however,
conserved to better than $5\times 10^{-5}$ of the initial value for
these tests at a value $(\gamma (\gamma-1))^{-1}\approx 1.78$. So the
kinetic energy growth in the particle distribution only corresponds to
about less than one in one thousand of the total. I.e. the kinetic
energy the particles obtain is taken from a slightly decreasing
internal energy allowing the total to be conserved to high precision.
The lower the neighbour number the faster that growth. The maximum
noise reached is controlled by the artificial viscosity. The noise
also decreases only very slowly over time after reaching the
maximum. This is one of the main reasons why particle settling is so
important in SPH simulations. The slow decline also shows why in
general settling procedures can be computationally quite intensive. In
the same figure we also plot results using {\em rpSPH} which
dramatically reduces this spurious kinetic energy creation keeping it
at zero to machine precision. 
%The efforts of
%\cite{2009MNRAS.395.2373C} amply demonstrate how dramatic SPH results
%depend on the initial particle distribution. 
\cite{2000PThPS.138..609R} caution that it makes no sense in standard
SPH to increase particle numbers while keeping the number of neighbors
fixed. Once a neighbour number is reached that keeps noise in the
force calculation to a minimum we find {\em rpSPH} to be stable while
only increasing the particle number. Note that we also have ran these
tests dramatically reducing the Courant factor without any improvement
in the case of standard SPH.

The thick solid line in Figure~\ref{fig:ke-noise} uses 20 neighbors
which seems optimal for this 2D calculation with the cubic spline
kernel. Here one has enough neighbors to estimate the gradients more
accurately while still having too few neighbours to show its pairing
instability. So one may be tempted to dismiss the finding that one has
the large velocity noise as long as one uses the ``correct'' number of
neighbours in one simulation. Unfortunately, this best choice,
however, is only applicable at the uniform density. To show this we
perturb the $x$ positions by a small amount so that the initially
uniform $x_0$ positions are changed by adding $\sin(2\pi\,x_0)/25$ to
them which gives central densities that are about 30\% above the
mean. We keep again the pressure to be exactly constant by setting the
entropy of the gas only once the density has been estimated from
kernel smoothing. The thick long dashed line in
Figure~\ref{fig:ke-noise} gives the associated velocity noise. It
again is of order one percent of the sound speed and grew very
rapidly.

One may also be tempted to to dismiss this particle noise as
irrelevant as it only contains less than a tenth of a percent of the
total energy of the system. However, we will see in the following it
is what leads to unphysical shear viscosity once one considers shear
flows further below.

\subsection{Absence of the Pairing Instability}

The velocity noise we just discussed is unfortunately not isotropic
nor is it random. It has a dominant component for velocities towards
directions of other particles and is an effect that aids the pairing
instability. Our reasoning here is somewhat contrary to the
explanations in the literature as e.g. in \cite{1981A&A....97..373S},
\cite{2008IJNMF..56...37V} or \cite{2009arXiv0906.0774R}. These
studies suggest that it is the shape of the kernel function that
causes clumping instability. 

Given that {\em rpSPH} does not show any spurious velocities in the
uniform density test given above while standard SPH develops clumps
within a few sound crossing times already shows that it cannot be the
shape of the kernel alone that is relevant here. In the following
tests we have looked for any sign of the clumping instability but have
not found any evidence for it independent of the number of neighbors
we used. This fact is certainly one of the factors that makes our
formulation have radically reduced errors in general. 

The clumping instability stems from particles pushing each other
closer to other particles. With a smaller distance to the other
particles the gradient of the kernel becomes smaller and in the next
time step the particle gets pushed further away from its initial
position. This way particles can pile up in the flat central part of
the smoothing kernel. That {\em rpSPH} does not show the clumping
instability makes one hopeful that even higher order kernels could now
be used to further improve the accuracy of the obtained solution.

\subsection{Dramatically Reduced Numerical Shear Viscosity}\label{sec:shear}

An easy two dimensional setup uses an adiabatic index of $\gamma=1.4$
in a unit square domain $x\in \{0,1\}$, $y\in \{0,1\}$ with periodic
boundary conditions. Particles are initialised on an exactly square
lattice with a density of $\rho(x,y)=1$ so that the initial density
estimate from the SPH kernel gives a density estimate of unity
to better than four parts in one thousand. We then add a sinusoidal
velocity perturbations to this uniform distribution. We set the
pressure to $P_0=\rho/\gamma$ to have a sound speed of unity. For the
first tests here we only use $50^2$ particles as there are no features
to resolve. In all cases we evolve to time $t=4$.

We choose $v_x(y) = \delta v_{\rm y}\cos(2\pi\, y)$ with $\delta
v_{\rm y}=1/2$. This shear flow setup gives an average kinetic energy
of $1/8$. A detailed discussion of how SPH behaves on this test for
different neighbor numbers, particle numbers and viscosity
prescriptions is given in the Appendix. In summary, it does very
poorly and transfers kinetic energy into heat very rapidly loosing
tens of percent in only four sound crossing times (two crossing times
of the fastest particles). It also gives {\em more} dissipation when
using more particles which makes a convergence study at fixed
neighbour number impossible. Below, when we discuss {\em rpSPH} for
viscous flow, we show that the effective numerical viscosity of
standard SPH is non-Newtonian and very large which explains why
standard SPH is inadequate to modeling fluids in general.

\begin{figure}
\includegraphics[width=0.48\textwidth]{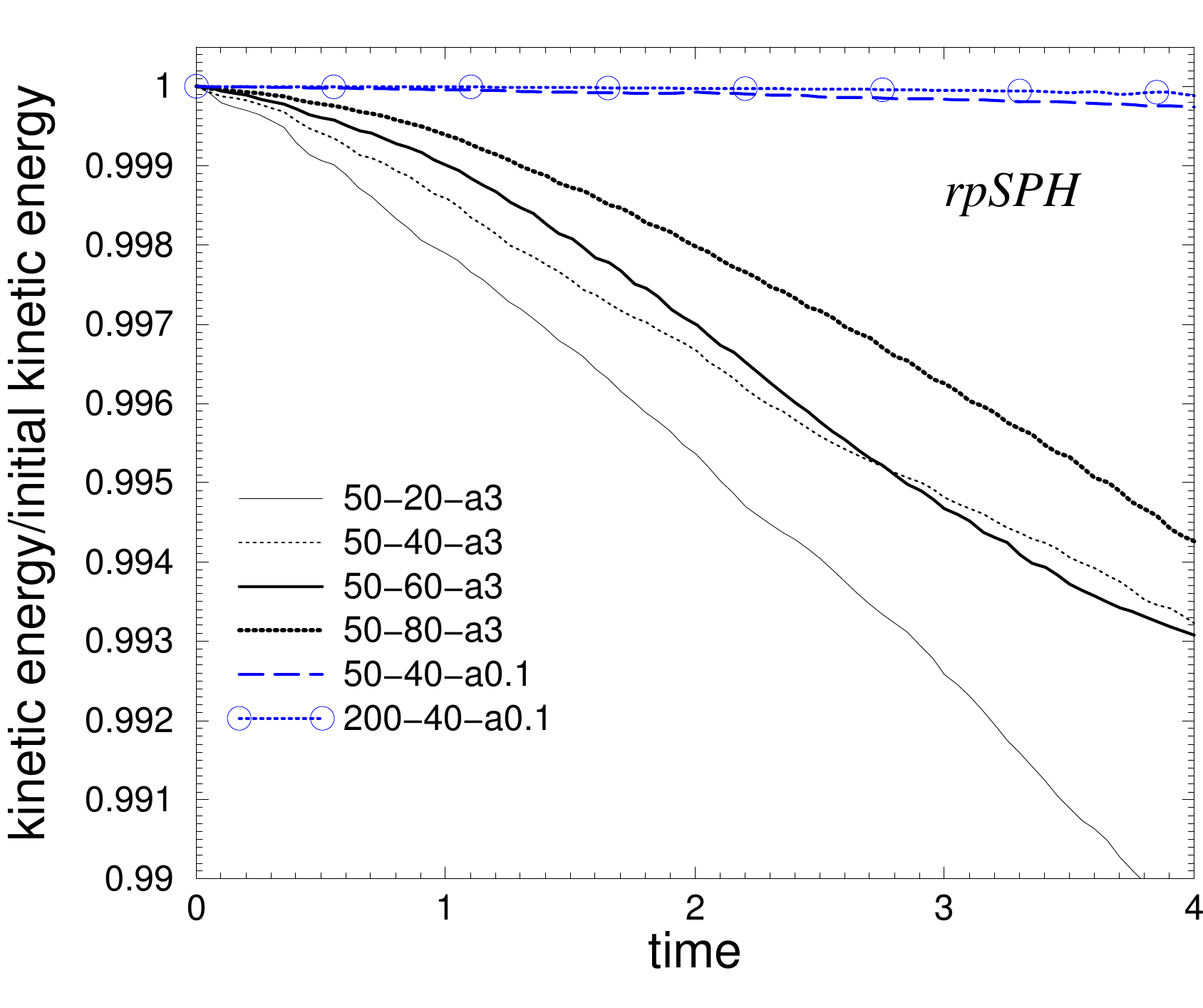}
\includegraphics[width=0.48\textwidth]{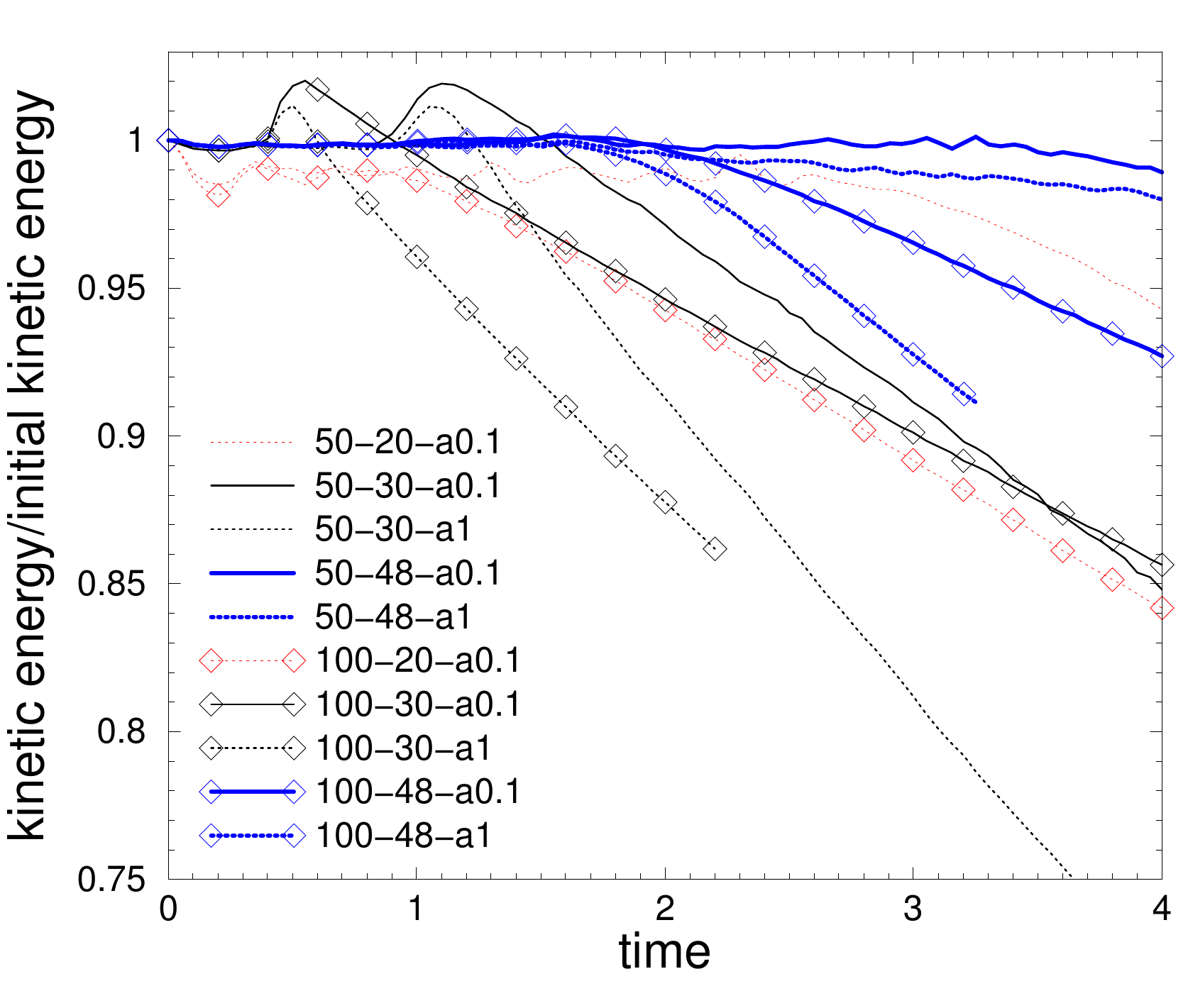}
\caption{The fraction of the total kinetic energy in the uniform
  density shear test as compared to its initial value as a function of
  time for {\em rpSPH}. The labels give the square root of the number
  of particles used, the number of neighbors and the artificial
  viscosity parameter. So the 50-40-a0.1 used $50^2$ particles, with
  40 neighbors and an artificial viscosity parameter of
  $\alpha=0.1$. We clearly see that using more neighbors always leads
  to less artificial shear viscosity. With more particles the
  effective numerical dissipation (which is very small to begin with)
  decreases.  The corresponding panel for SPH below has a $y$-axis 30
  times as big.  (Bottom) The fraction of the total kinetic energy in
  the uniform density shear test as compared to its initial value as a
  function of time for standard SPH.  We clearly see that using more
  neighbors always leads to less artificial shear viscosity. However,
  increasing the particle number leads to {\em more}
  dissipation. }\label{fig:reduced-S}
\end{figure}

The results for {\em rpSPH} are summarized in
Figure~\ref{fig:reduced-S} which is to be compared to the bottom panel
for SPH, plotting the kinetic energy in the box as a function of sound
crossing times. Note that the $y$-axis in the two panels of
Figure~\ref{fig:reduced-S} is different by a factor of 30.

We should note that this test problem when used in typical Cartesian
grid codes will give zero numerical dissipation to machine precision
because the uniform nature of the flow along the grid axes.

\cite{PriceThesis2004} considered seemingly similar tests in his
thesis. However, note that there an isothermal equation of state was
used. He shows one case with $c_s=0$, i.e. a pressure less fluid and
another case with $c_s=0.05$. His initial shear profile is like the
one we choose here but with twice the amplitude so the fastest
particles move with unit speed. The pressure-less case is irrelevant
here since without pressure forces the discretization of the momentum
equation cannot make a difference. For the second case with $c_s=0.05$
he gives the result only after one sound crossing time at $t=20$. The
density fluctuations have grown to order of a percent of the initial
value after that single sound crossing time.  As
Figure~\ref{fig:reduced-S} shows for most reasonable choices of
neighbor numbers not much kinetic energy is dissipated over this time
also in our simulations with an adiabatic index of $1.4$. We also
repeated this isothermal shear test and find results consistent with
\cite{PriceThesis2004}.  This again emphasises that as long one is
interested in very few sound crossing times, SPH can give correct
results and gives an indication that this is not specific to Gadget. 

This problem also allows us to measure the effective Reynolds numbers
one can hope to model with standard SPH. 

Solving analytically the incompressible Navier-Stokes
equations for our initial conditions because to good approximation
only the viscous term is relevant we have
$$
{\partial v \over \partial t} = \nu \nabla^2 v.
$$
Since the variables are separated we can easily find that
\begin{equation}
  \label{equ:sol}
  v(x,t) = e^{-4\pi^2\nu\, t} \, v(x,t=0),
\end{equation}
showing that the initial shape of the profiles should not change.  We
can now use equation~\ref{equ:sol} to get an estimate of the kinematic
viscosity from the fraction of kinetic energy remaining 
\begin{equation}
  \label{eq:kin}
  \nu = { \ln(1/F) \over 8\pi^2 t},
\end{equation}
where $F$ denotes the fraction of the kinetic energy remaining up to
time, $t$. This is $\nu\approx 3.2 \times 10^{-3} \ln(1/F)$ at four
crossing times. 

 The Reynolds number measures the ratio of inertial forces,
$\rho v^2$, to viscous ones, $\mu V/L$, where $L$ is the
characteristic length scale, $V$ the mean velocity and $\mu$ is the
dynamic viscosity. So $R=V L/\nu$ with $\nu = \mu/\rho$ the kinematic
viscosity.  Despite the ambiguities we may take $L=1/4$ the quarter
wavelength of the velocity perturbation, $V=\sqrt{N_P^{-1}\sum_{N_p}
  v^2}=1/2\sqrt{2}$ the root mean squared velocity.

So for the typical values of $F$ we found for SPH say 70 (97) per cent
remaining equation~(\ref{eq:kin}) gives a kinematic viscosity of $\sim
10^{-3}$ ($10^{-4}$). Consequently the numerical Reynolds number
$R=LV/\nu\sim (8 \sqrt{2} \, \nu)^{-1}\sim 90/(1000 \nu)$.  This is
very low and lower than most observed transitions between laminar and
turbulent flows in the laboratory or terrestrial applications.

The analytic solution in equation~(\ref{equ:sol}) only holds if the
fluid is Newtonian so that the shear stress can be described by the
single number of the kinematic viscosity $\nu$. Because,
Figure~\ref{fig:uni}, shows that the initial velocity profile changed
strongly one also concludes that the fluid flow as modeled by SPH is
that of a non-Newtonian fluid. So while it would have been interesting
to think of SPH as solving effectively a the Navier Stokes rather than
the Euler equations we see that this is not the case. The effective
numerical viscosity is non-Newtonian and does not in general decrease
with numerical resolution.

\subsection{Kelvin--Helmholtz Instability}
The Kelvin--Helmholtz instability occurs at the interface of two
shearing fluids of different densities when velocity perturbations
perpendicular to the interface grow to eventually mix the layers. In
the inviscid case this is understood analytically
\citep{1961hhs..book.....C} and the growth time scale for a sinusoidal
mode of wavelength $\lambda$ between two fluids of density $\rho_1$
and $\rho_2$ with a shear velocity $v=v_2-v_1$ between them is
\begin{equation}
\label{eq:KHItime}
\tau_{\rm
  KH}\equiv\frac{2\pi}{w}=\frac{(\rho_1+\rho_2)\lambda}{(\rho_1\rho_2)^{1/2}v}.
\end{equation}
This problem is typically setup with an adiabatic index of
$\gamma=5/3$ in a unit square domain $x\in \{0,1\}$, $y\in \{0,1\}$
with \citep[e.g.][]{2009arXiv0906.0774R}
\begin{equation}
\rho,T,v_x=\left\{
 \begin{array}{rl} \label{equ:KHsetsharp}
 	\rho_1,T_1,v_1  & |y-0.5|< 0.25\\
 	\rho_2,T_2,v_2  & |y-0.5|> 0.25
 \end{array} \right.
\end{equation}
We choose $\rho_1=2, 5, 10$ and $\rho_2=1$ and the uniform pressure
$\rho_2/\gamma=3/5$ which gives a sound speed of $1$ in the low
density surrounding medium. This standard setup then perturbs the
interface with
\begin{eqnarray}
v_y & =& \delta v_{\rm y}[\sin(2\pi
(x+\lambda/2)/\lambda)\exp(-(10(y-0.25))^2) \nonumber \\
& & + \sin(2\pi x/\lambda)\exp(-(10(y-0.75))^2)]
\end{eqnarray}
where we choose $\lambda=1$ and vary $\delta v_{\rm y}$.  So for our
density contrasts the growth times are $\tau_{\rm KH} \approx 2.12$,
2.68 and 3.49 for the density contrasts 1:2, 1:5, and 1:10,
respectively.

It would seem prudent to compare these results to the many
investigations that recently have addressed the KH instability using
SPH \citep{2007MNRAS.380..963A, 2008MNRAS.387..427W,
  2008JCoPh.22710040P, 2009arXiv0906.0774R,
  2009arXiv0912.0629H}. However, all of them chose a setup that
\cite{2009arXiv0909.0513R} showed to be ill defined. While all
these studies where concerned with the issue of whether SPH can handle
KH instabilities at all, they do not ask whether it actually converges
to a correct solution. This can be seen e.g.  in
\cite{2009arXiv0906.0774R} where their modified SPH solution compares
already visually poorly to the corresponding Eulerian result.

Following \cite{2009arXiv0909.0513R} we opt for a more well defined
setup for which they explicitly showed convergence. We modify the
initial density, and velocity profile using the ``ramp'' function
\begin{equation}
\label{equation:ramp}
R(y) = \frac{1}{1 + \exp[ 2(y-0.25)/\Delta_{y}]}\frac{1}{1 + \exp[ 2(0.75-y)/\Delta_{y}]},
\end{equation}
choosing $\Delta_{y} = 0.05$ so that $\rho(y) = \rho_{2} +
R(y)[\rho_{1} - \rho_{2}]$ and the velocity shear is $v_{x}(y) = v_{2}
+ R(y)[v_{1} - v_{2}]$. For the initial velocity perturbation we take
$v_{y}(x) = \delta v_{\rm y} \sin(n\pi x)$ setting $\delta
v_{\rm y}=0.1,\, \ 0.01$ and $n=2$.

\begin{figure*}
\centerline{
\includegraphics[height=10.3cm,clip=true,trim=1cm 0cm 0.2cm 0]{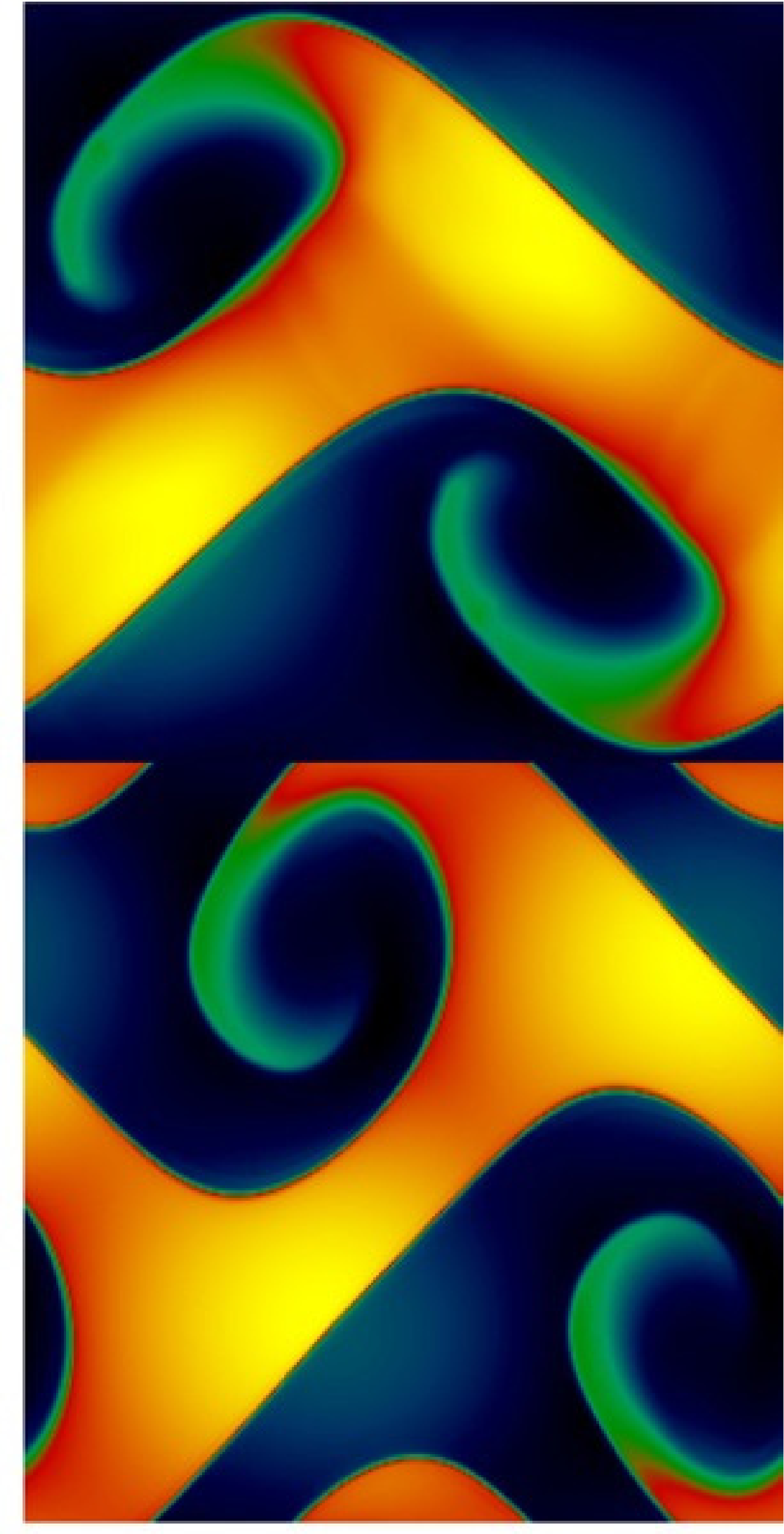}
\hspace{-0.1cm}
\includegraphics[height=10.7cm,clip=true,trim=0cm 1cm 5cm 0cm]{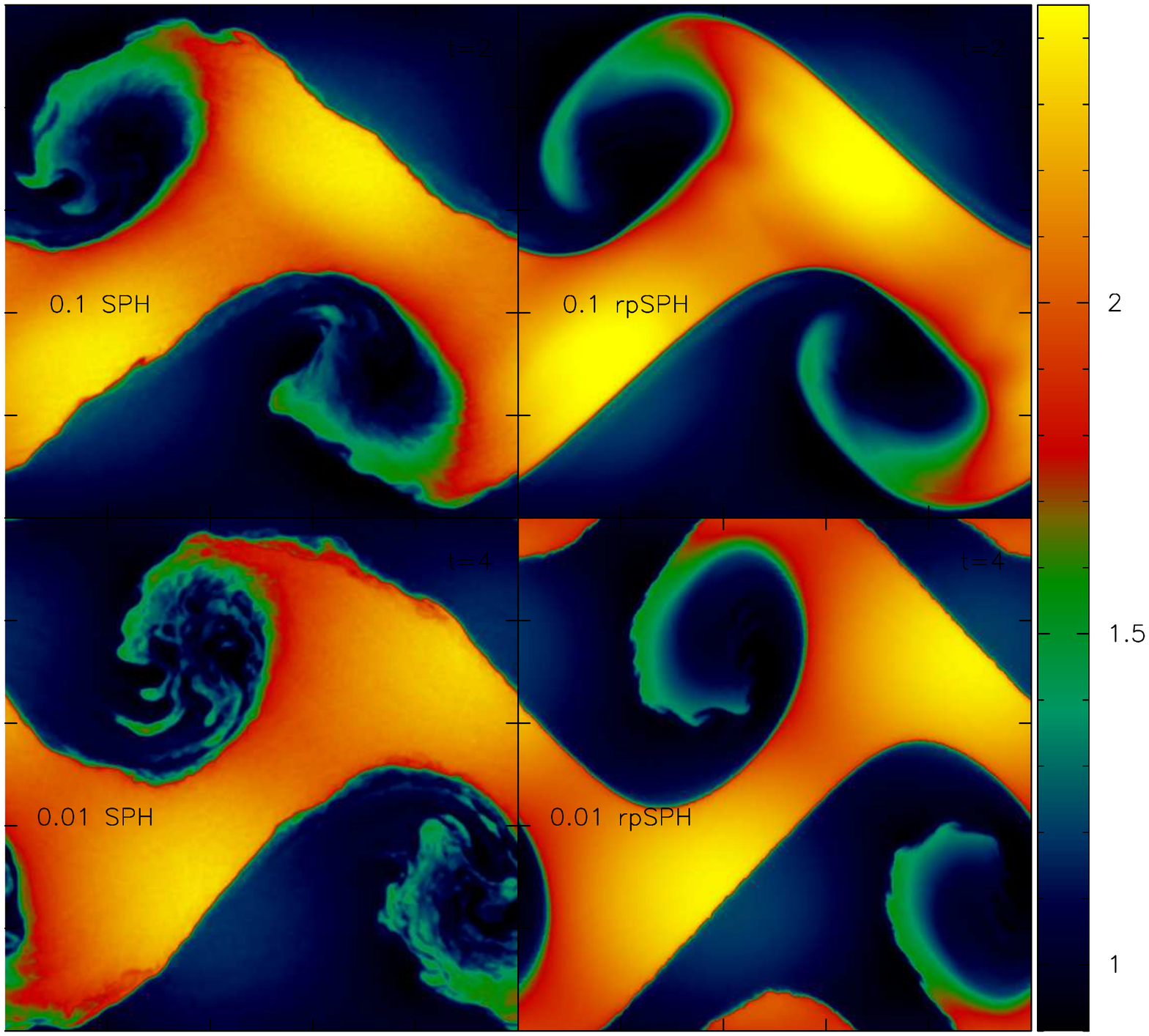}
}
%\vspace{-12.cm}
\caption{Comparison of the final density distribution in a two
  dimensional Kelvin--Helmholtz test. The velocity perturbation is a
  tenth of the sounds speed (0.1) in the top panels and one hundredth
  (0.01) in the bottom row. The left simulations show a grid based
  solution with $256^2$ using the piecewise parabolic method. The
  middle row gives results for standard SPH and on the right are {\em
    rpSPH} calculations. This problem is set up carefully to not have
  any physical small scale structure and has been shown to converge
  with grid codes using a resolution exceeding $2500^2$
  \citep{2009arXiv0909.0513R}. {\em rpSPH} behaves significantly
  better in not breaking up from discretization noise. The bottom left
  shows how SPH fails to grow at the correct rate and is generally
  more noisy. All the particle based simulations used 1000$^2$
  particles and 100 neighbors with $\alpha=3$.  }\label{fig:KH-comp}
\end{figure*}

Figure~\ref{fig:KH-comp} compares results obtained with {\em Enzo}
\citep{1997ASPC..123..363B, Bryan:2001, 2004astro.ph..3044O}, standard
SPH and {\em rpSPH} for the two different initial velocity
perturbations. The improvement of {\em rpSPH} over SPH is
dramatic. The billows grow at the correct rate and show a minimum of
artificial small scale structure. The figure shows the highest resolution
we have calculated. However, even with 120$^2$ particles 
neighbours one can obtain correct results using {\em rpSPH}. 
Also our choice of a high $\alpha$ is inconsequential in {\em rpSPH}
in this incompressible setup since the Balsara switch reduces it
dramatically in practice. We left this high value just to show
that one can get an accurate solution without having to adjust his
viscosity parameter.

\subsection{Rayleigh Taylor instability instabilities}
\begin{figure}
\centerline{
\includegraphics[width=0.4\textwidth]{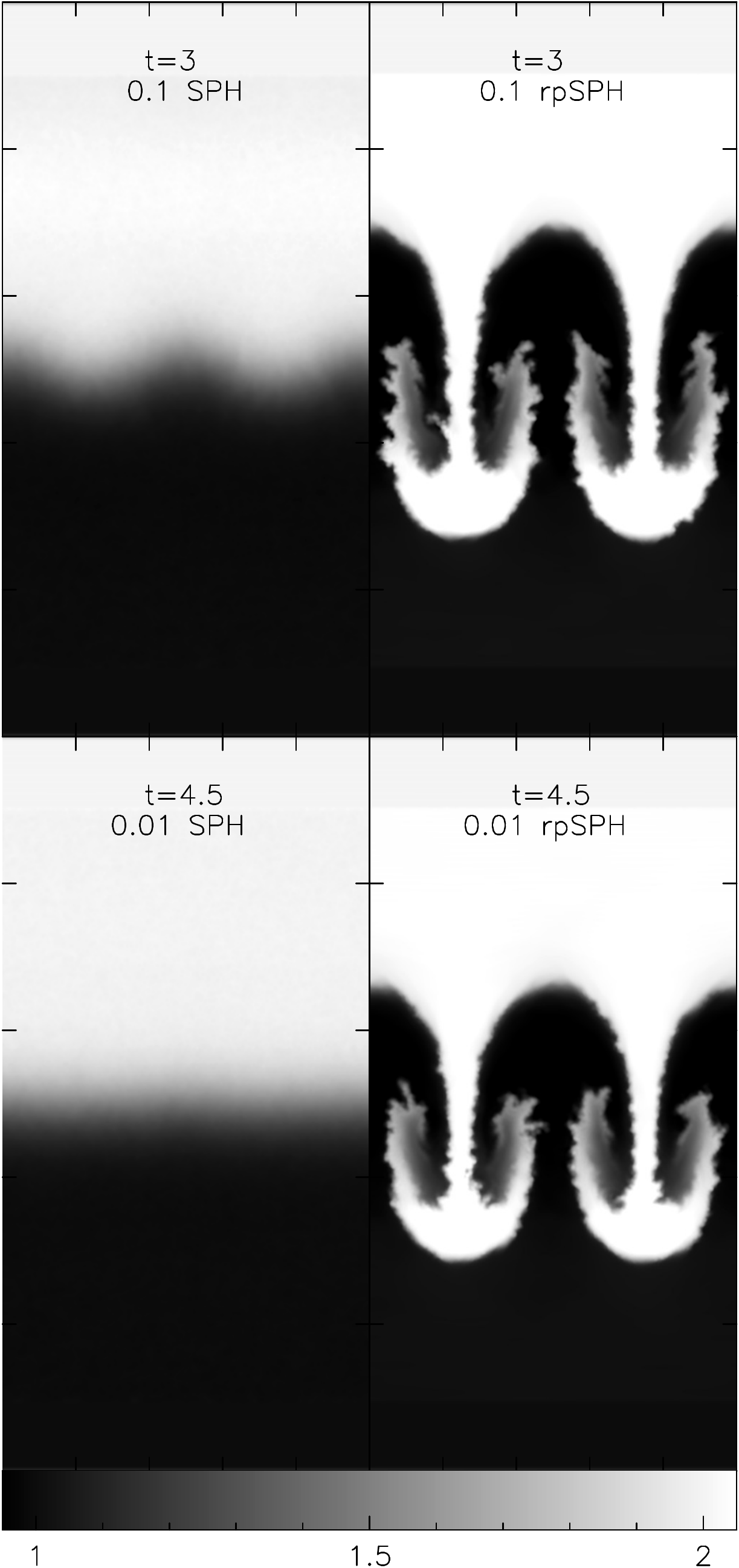}
}
\caption{Comparison of the final density distribution in a two
  dimensional Rayleigh Taylor test between standard SPH (left panels) and {\em
    rpSPH} (right panels) for two different initial velocity
  perturbations $\delta v=0.1$ (top panels) and $\delta v=0.01$
  (bottom panels). The unphysical ``surface tension'' of SPH prevents
  the growth of the instability entirely. {\em rpSPH} easily recovers
  the correct behaviour. All simulations here used $500^2$ particles
  and 70 neighbours.}\label{fig:rpSPH-RT}
\end{figure}

Another classic test of a code's ability to handle subsonic
perturbations is the RT instability
\citep[e.g][]{2000ApJS..131..273F,2007ApJ...671.1726S,2008ApJS..178..137S}.
SPH has been used to model supernova explosion previously and RT
instabilities have been seen \citep{1994MmSAI..65.1013H} as well as
global convective instabilities \citep{2004ApJ...601L.175F}.  The
physical situation \citep{1961hhs..book.....C} here consists of a
heavy fluid being supported by a lighter fluid against which initially
are in pressure equilibrium with a constant acceleration
(e.g. gravity).  Remarkably, all idealised test cases that we are
aware of use an initially unresolved contact discontinuity and
consequently no converged results independent of the method of
solution have so far been presented. Instead the differences between
different reconstructions schemes, Riemann solvers, meshing, etc. all
contribute to the final structures produced in these simulations.

Similarly as the Kelvin--Helmholtz problem above, we chose here initial
conditions that try to minimize the computational requirements while
yielding converged results. The two dimensional domain is setup with
periodic boundary conditions along the $x$ direction with $x\in
\{0,1/2\}$ and reflecting boundary conditions along $y$ with $y\in
\{0.1,0.9\}$. To achieve the reflecting boundary conditions in
Gadget-2 we set up the density distribution in $y\in \{0,1\}$ and make
all particles with $y<0.1$ and $y>0.9$ stationary ({\tt
  -DSPH\_BND\_PARTICLES} compile option). These particles are not
allowed to change their entropy or positions consequently they retain
their initial pressure and density. Particles that are trying to
penetrate through these walls have their positions changed to be at
the wall and $y$ velocity vectors reversed. This is only a crude way
of modeling reflecting boundaries with SPH but will suffice here to
compare between SPH and rpSPH.

In keeping with past literature we use an adiabatic index of
$\gamma=1.4$, and set up the density at the top to be $\rho_1=2$ and
$\rho_2=1$ at the bottom.  So the density profile is $\rho(y) =
\rho_2+ (\rho_1-\rho2)/[1+\exp(-2(y-0.5)/\Delta_{y})]$ with
$\Delta_{y}=0.05$ in the cases presented here. The velocity
perturbation is applied in $y$ direction with $v_{y}(x,y)=\delta
v_{\rm y} (1+\cos(8\pi(x+1/4))) (1+\cos(2/0.4\, \pi(y-1/2)))/4$ and
the $y$ velocities are set to zero for $y$ positions above 0.7 and
below 0.3.  The pressure is set to $P_0=\rho_1/\gamma=10/7$ to give a
sound speed of one at the interface and is set into hydrostatic
equilibrium with the constant acceleration $g=1/2$ in the negative $y$
direction with $P(y)=P_0-g \rho(y)(y-1/2)$. This gives a pressure
difference of 60\% between the top and the bottom of the domain.  The
smaller this pressure difference the more difficult it becomes for SPH
to model it.  Similarly, the initial velocity perturbation
again should not be very much smaller than the sound speed in order to
survive viscous damping before a growth time of the instability.

We present the results for this test for velocity perturbations of
$\delta v=0.1$ and $\delta v=0.01$ in Figure~\ref{fig:rpSPH-RT}. 

The differences are dramatic. Where SPH fails completely to see growth
of the instability {\em rpSPH} gives the expected behaviour for both
perturbation strengths. 

\begin{figure}
\includegraphics[width=0.65\textwidth]{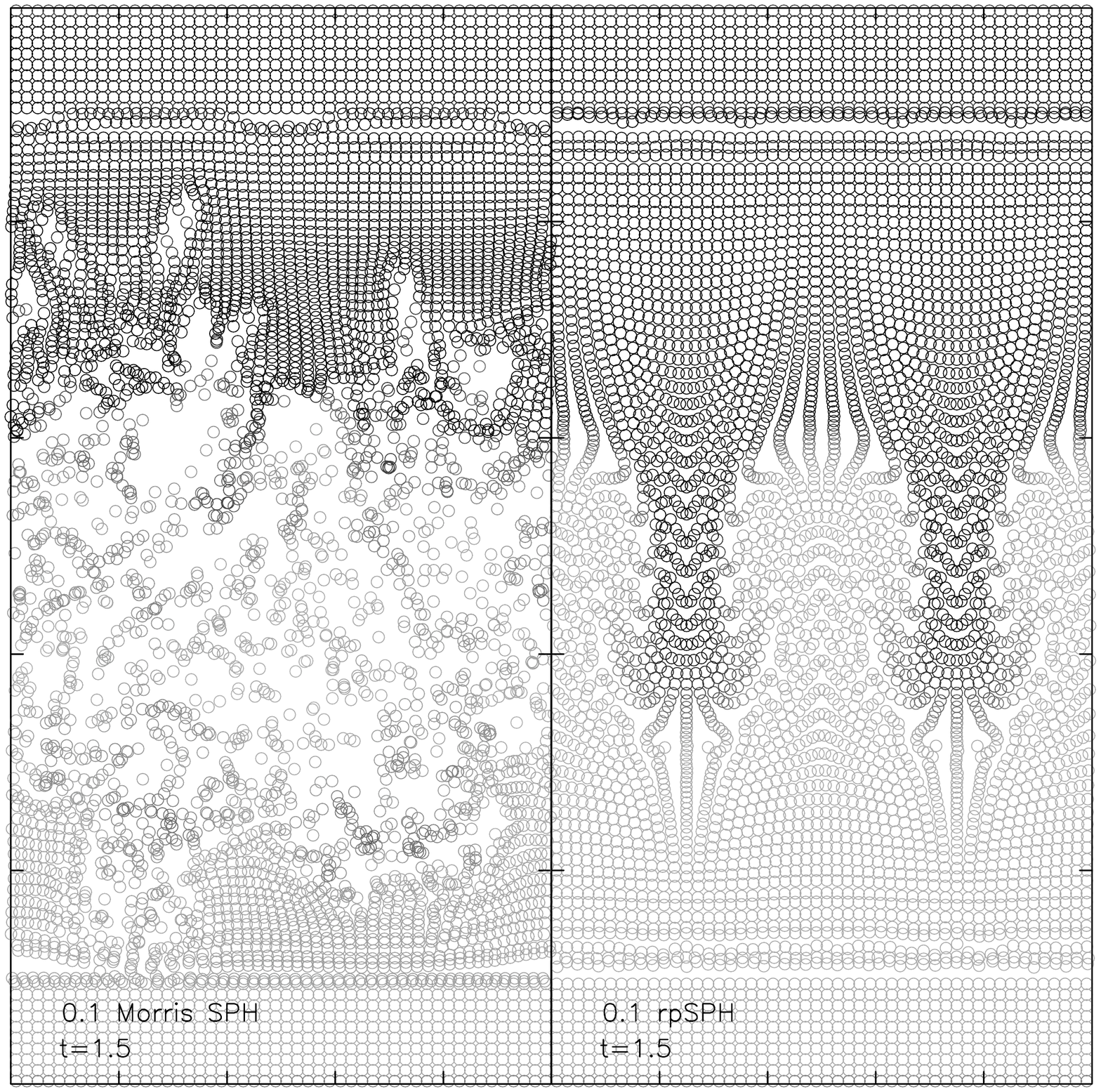}
\caption{Density in a Rayleigh--Taylor test with a ten times heavier
  fluid on top than the bottom. Left is the Morris formulation and on
  the right is {\em rpSPH}. Both Morris SPH and {\em rpSPH} runs used an
  initial uniform grid of 50 by 100 particles with varying masses to
  describe the higher density for the interface and top fluid. The
  Morris formulation is unstable in this problem while {\em rpSPH}
  behaves as expected. }\label{fig:mSPH-RT}
\end{figure}

That {\em rpSPH} is dramatic improvement over Morris' formulation
despite only differing in one index is seen in
Figure~\ref{fig:mSPH-RT}. There we give a Rayleigh Taylor problem at
low resolution of $100x50$ particles and a density ratio of $10$ as
further discussed in section~\ref{sec:mm}. All parameters were the
same. A courant factor of $0.2$ is used, a neighbour number of 40,
$\alpha=1.5$, Balsara switch is on, and the initial velocity
perturbation amplitude is $0.1$. Clearly our formulation is more
stable lending support to our discussion on the different error
properties of the two discretisations given above.

\subsection{Shock tubes}

\subsubsection{Sod shock tube}
\begin{figure*}
\includegraphics[width=0.9\textwidth]{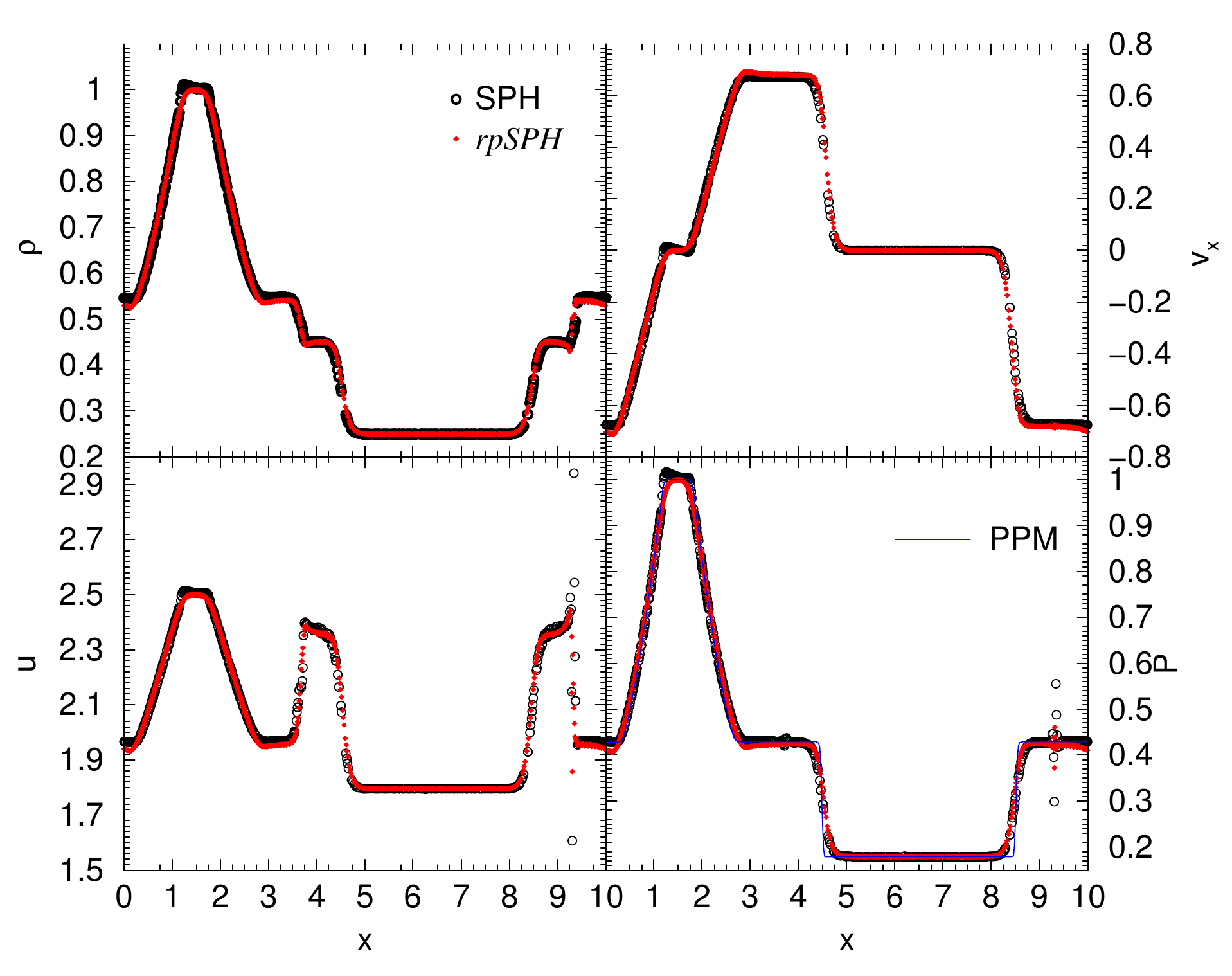}
\caption{Clockwise the density, $x$-velocity, pressure and specific
  internal energy are given for the classic Sod shock tube
  \citep{1978JCoPh..27....1S} at time $t=1$. The simulations used
  $200^2$ particles, $80$ neighbours with a tolerance of one,
  $\alpha=3$ for both the give the {\em rpSPH} solution (filled
  diamonds) and standard SPH (open circles). We plot every 50th
  particle of the 40,000 employed.  In the pressure panel (bottom
  right) we also show a references solution computed with a Eulerian
  code using the piecewise parabolic method. The agreement is quite good.
  {\em rpSPH} handles the contact discontinuities much better and the
  ``blip'' seen in SPH is absent in the one form the smoothed
  interface ($x\sim3.8$) while it is much reduced in the one that was
  initially sharp ($x\sim9.3$).}\label{fig:Sod-200}
\end{figure*}

So far we have tested our new formalism only in very weakly
compressible situations. We will use the classic Sod shock tube
\citep{1978JCoPh..27....1S} to compare {\em rpSPH} to standard SPH
here. \cite{2009NewAR..53...78R} recently, gave the results for
varying viscosity prescriptions and including artificial conduction
terms. We change the setup only slightly. The left state has a
density and pressure of unity while the right state has a quarter of
the density and a pressure of $0.1795$. This test is evolved with an
adiabatic index of $\gamma=1.4$ and we set it up as a two dimensional
problem with equal mass particles in a box that extends from zero to
ten in $x$ and zero to one in the $y$-direction. We choose 40 rows of
particles in the $y$ direction and vary the spacing along $x$ to
achieve the given densities using a total of $200^2$ particles which
are initially at rest. Additionally we set the interface to be at
$x=3$ and smooth it with an exponential ramp such that all
hydrodynamic variables are given by $ r + (1+\exp(2*(x-3)/\delta
x))^{-1}(l-r)$ where we take $\delta x = 0.05$ and $l$ and $r$ denote
the left and right states. We employ periodic boundary conditions
which gives us another interface at $x=10$ which has the reversed left
and right states but has an initially discontinuous state. This will
give us the opportunity to show the difference between smoothed
interfaces and artificially sharp to be visible in one figure. For the
first results we present we have used 80 neighbors and an artificial
viscosity parameter of $\alpha=3$ for both the SPH and the {\em rpSPH}
calculation. Both employ the Balsara switch to limit the viscosity
which will play no role here because there is no rotational component
to the flow. Figure~\ref{fig:Sod-200} summarizes the hydrodynamic
state variables at time $t=1$. To first approximation one gets
identical results with the new formalism as compared to the standard
approach. Linear momentum is not conserved in {\em rpSPH} and we find
a linear excess velocity of $(-105,-3\times 10^{-5})$ so per particle
an error on the velocity of $0.003$ in the $x$-direction and a
completely negligible component along the $y$-direction. This is at a
time when the r.m.s. velocity is $\sim 0.46$ so just slightly above
one half of a per cent error in the dominant $x$-velocity.  SPH has
poor behaviour at the contact discontinuities. For both the one
originating from the initially smoothed and the the discontinuous
interface at the right boundary. Both contacts at $x\sim3.8$ and
$x\sim 9.3$ are better captured by {\em rpSPH}.  The Sod shock tube
has few features and it is reassuring that using as many as $200^2$
particles can give an excellent answer.
There are only slight differences in how {\em rpSPH} handles one
dimensional shock tubes. We will discuss one very popular application
taken from a cosmological context after testing a very strong shock
next.

\subsubsection{Strong Shock}

Here we give another test of a much stronger shock than the one by
sod. This one has a Mach number close to one hundred. We also use the
chance to compare this to the difference formulation studied by 
\cite{1996PASA...13...97M}. The density and pressure are $(1,6.6\times
10^4)$ on the left and $(1/5,1)$ on the right. This is very similar to
the one studied by \cite{2006MNRAS.367..113P} and is well known to
work well with standard SPH. Here we use $35$ neighbours, $\alpha=4$
and 5000 particles. This is a good example where one can make {\em
  rpSPH} and the Morris formulation give unphysical results. These
methods require the pressure gradient to be resolved. So if you start
with completely discontinuous left right states one will get
unphysical waves giving unexpected results. However, this is not a
shortcoming of the method but simply are errors that come from not
resolving the initial conditions. We again use the ramp function from
above with a width of 4 in this very long domain ranging from 0 to 500
in $x$ and $0$ to $10$ in $y$. 
\begin{figure*}
\includegraphics[width=0.9\textwidth]{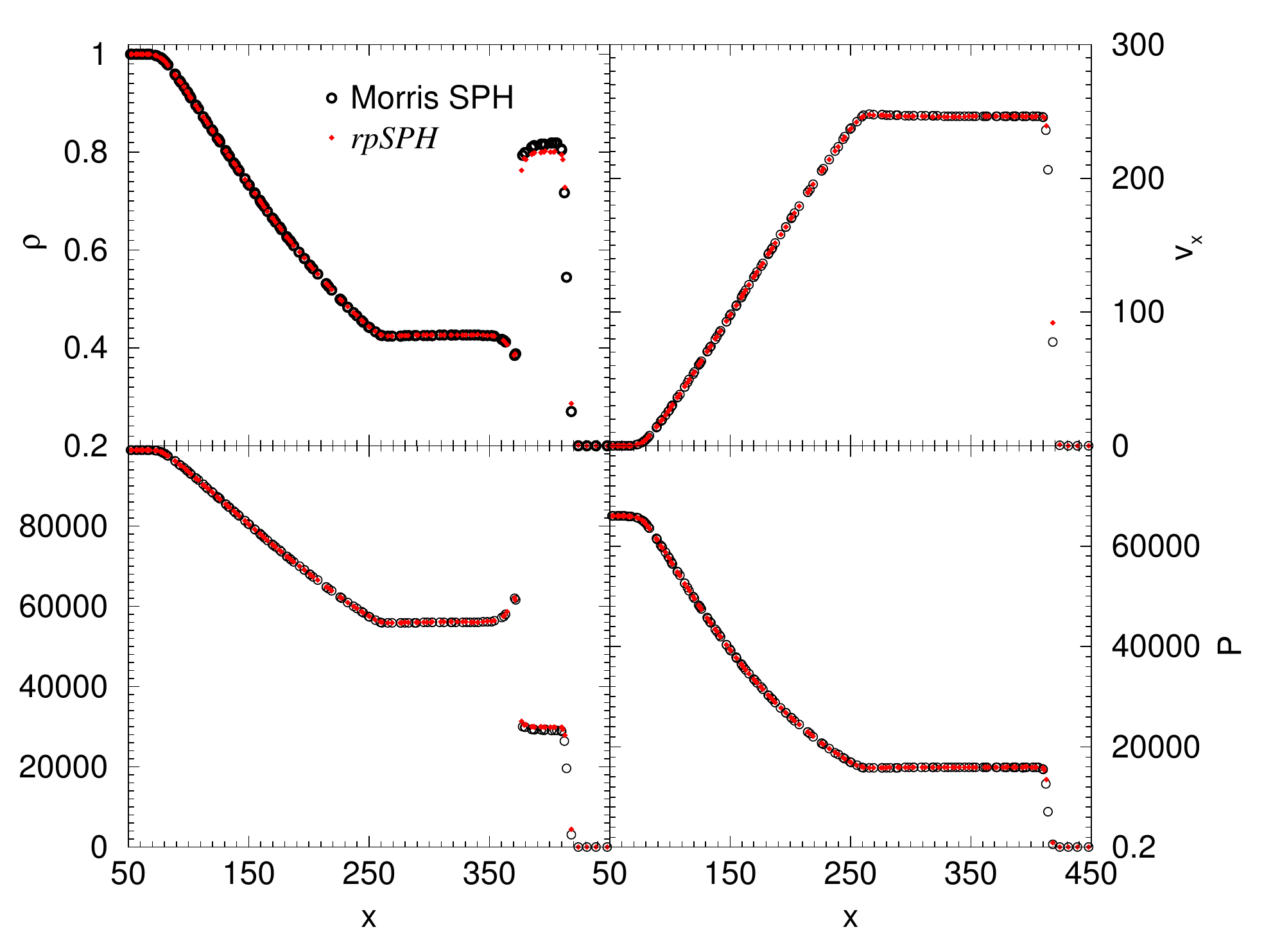}
\caption{Clockwise the density, $x$-velocity, pressure and specific
  internal energy are given for a strong shock with Mach number $\sim
  100$ at time $t=0.5$. We used The simulations used $5000$ particles
  arranged as ten rows in y in an elongated domain ${{0,500},{0,10}}$,
  $35$ neighbours with a tolerance of one, $\alpha=4$. We give the
  result obtained with Morris' formulation (open circles) and {\em
    rpSPH}. We plot every 50th particle of the 5,000 employed.  Both
  approaches handles the contact discontinuities much better and the
  ``blip'' seen in standard SPH is gone. {\em rpSPH} gets the correct
  density jump of a factor of 4 better than the Morris
  formulation.}\label{fig:StrongShock}
\end{figure*}

We cannot confirm Morris' claim that his formulation gives large post
shock oscillations in this method and suspect that he may have set up
discontinuous initial conditions.  We can see that our new formulation
performs somewhat better than the Morris formulation as it does not
overshoot the analytical density jump of 4 raising the density from
$0.2$ to $0.8$ in Figure~\ref{fig:StrongShock}. Otherwise both
approaches work fine and have no problem in modeling strong shocks and
evolving it for large distances.

\subsection{Sedov--Taylor Blast Wave}

\begin{figure}
\includegraphics[width=0.45\textwidth]{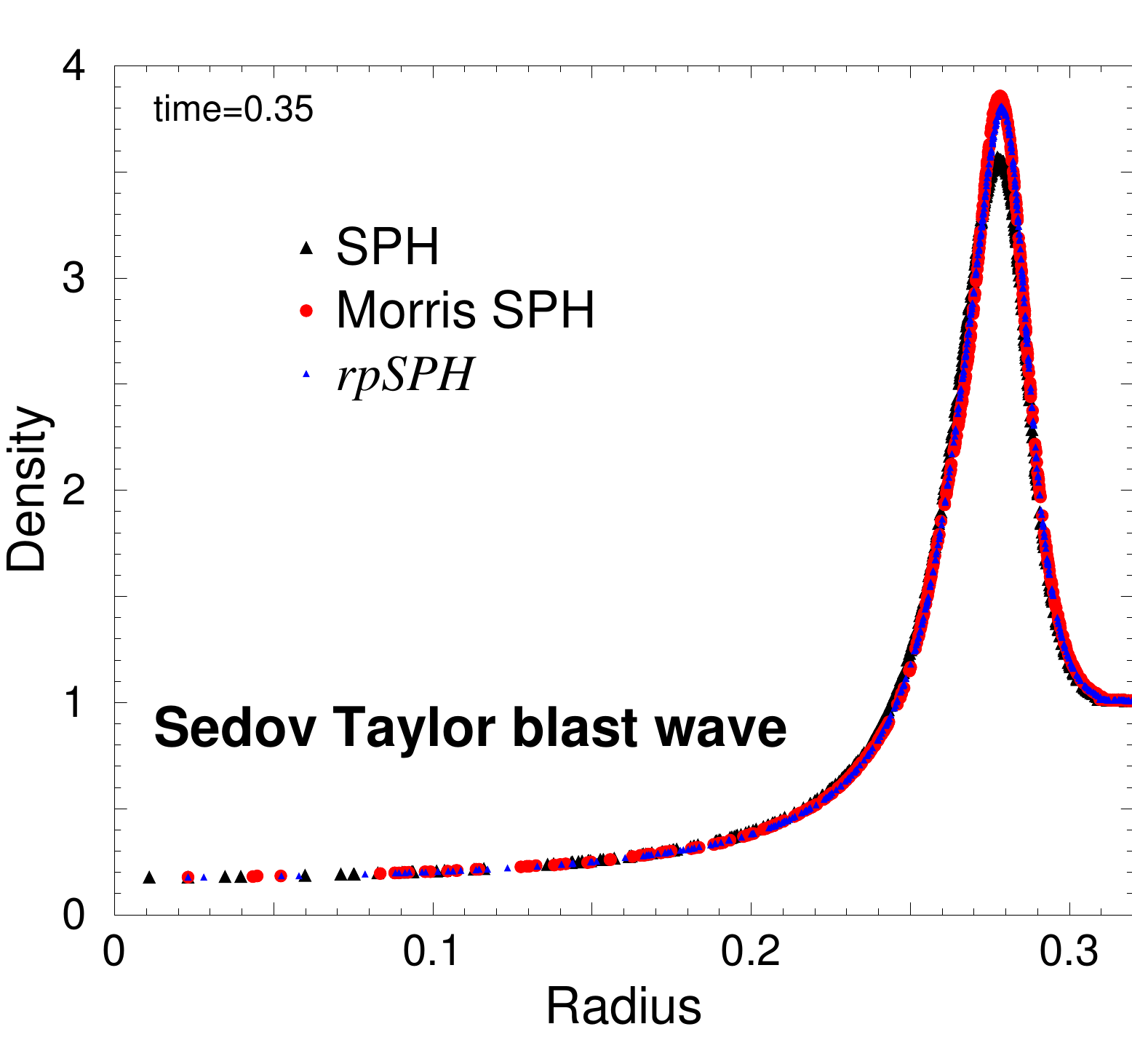}
\caption{Comparison of the shock profile in the density between SPH
  and {\em rpSPH} simulations. The analytic solutions demands a
  density jump of four which {\em rpSPH} and the Morris formulation
  stay close to over the course of the calculation. Standard SPH falls
  a little short, however, it is insensitive to choices of initial
  conditions and is a better choice if one cannot afford sufficient
  resolution to model the exploding region.
}\label{fig:Sedov}
\end{figure}

Another particularly strong shock is formed in the Sedov-Taylor blast
wave \citep{1959flme.book.....L} presenting a difficult test problem
for incompressible hydrodynamics codes. One the one hand it is a
self similar solution which makes it insensitive to how exactly one
sets it up as long as one evolves the system for a very long time. On
the other hand it is the solution for a point explosion. For a given
finite resolution, however, there is no unique way of specifying the
initial conditions. Here is where exact momentum and energy
conservation is very helpful as one can set up the initial conditions
at will and even if one were to make very large errors in the time
evolution the method will still arrive at the self similar
solution. In conservative grid codes this still can lead to aspherical
solutions if one did not resolve the spherical central hot region. 
Since SPH however uses spherical kernels one can get away sometimes
even by just heating one single particle
\citep{2002MNRAS.333..649S}. This is very useful in applications such
as galaxy formation simulations where one is always far from resolving
the relevant length scales of an explosion. On the other hand any
physics that were to occur at a scale of the shell thickness would be
impossible to resolve in such a single particle energy ejection. 
For {\em rpSPH} and the Morris formulation we need to resolve the
pressure gradients in the initial conditions as we saw above in the
strong shock setup. 

We set up a square lattice of particles with 300 particles on a side
in the unit domain. For resolved initial conditions we set a spherical
region in the center of radius $r=0.1$ with the same ramp function as
above using a width of $0.1$, to have a sound speed of one for an
adiabatic index of $\gamma=5/3$. For both simulations we used a
Courant number of 0.2 ($0.1$ in Gadget), 80 neighbors, artificial
viscosity $\alpha=2.5$ and had the Balsara switch
off. Figure~\ref{fig:Sedov} shows that there potentially is also an
advantage to {\em rpSPH} simulations when modeling shocks. We have
failed to get standard SPH to give a stable correct density jump of
$(\gamma+1)/(\gamma-1)=4$ for our setup. Also the three dimensional
versions shown by \cite{2002MNRAS.333..649S} always seem to be too low
by as much as a factor of two. However, standard SPH is much less
sensitive to how one sets up the initial conditions and performs much
better at low resolutions.

% In this problem we notice that to arrive at the correct density jump
% one should use the arithmetic mean of the kernel derivatives in
% equation~(\ref{eqnmotrpSPH}) if one also uses that mean when
% calculating the viscous forces. It is also possible to use the
% gradient of the kernel computed by particle $i$ to compute the viscous
% force while using the one of particle $j$ to compute the pressure
% force which makes physical sense but can lead to noise in almost
% incompressible flows.

These strong shock problems can work but clearly are not the biggest
strengths of {\em rpSPH}. However, at this point we simply reused the
artificial viscosity prescription which was designed for standard
SPH. We believe it is likely that one can find an alternative
formulation for the artificial viscosity that fits better into the
{\em rpSPH} discretisation which may improve its behaviour for highly
supersonic conditions. 
Until a better artificial viscosity prescription is designed one may
opt to switch between standard SPH and {\em rpSPH} based on the local
divergence. We have successfully applied this strategy by using a
switch that evaluates the standard SPH sum if $-h_i \div \vec{v}_i > 3
c_{s,i}$ and the {\em rpSPH} sum for less strongly convergent flow. 
Here $h_i$ and $c_{s,i}$ denote the smoothing length and the current
sound speed of the particle. This formulation is robust in all our
tests.

\subsection{Cosmological Integration of the Santa Barbara Galaxy
  Cluster}

\begin{figure*}
\includegraphics[width=0.95\textwidth]{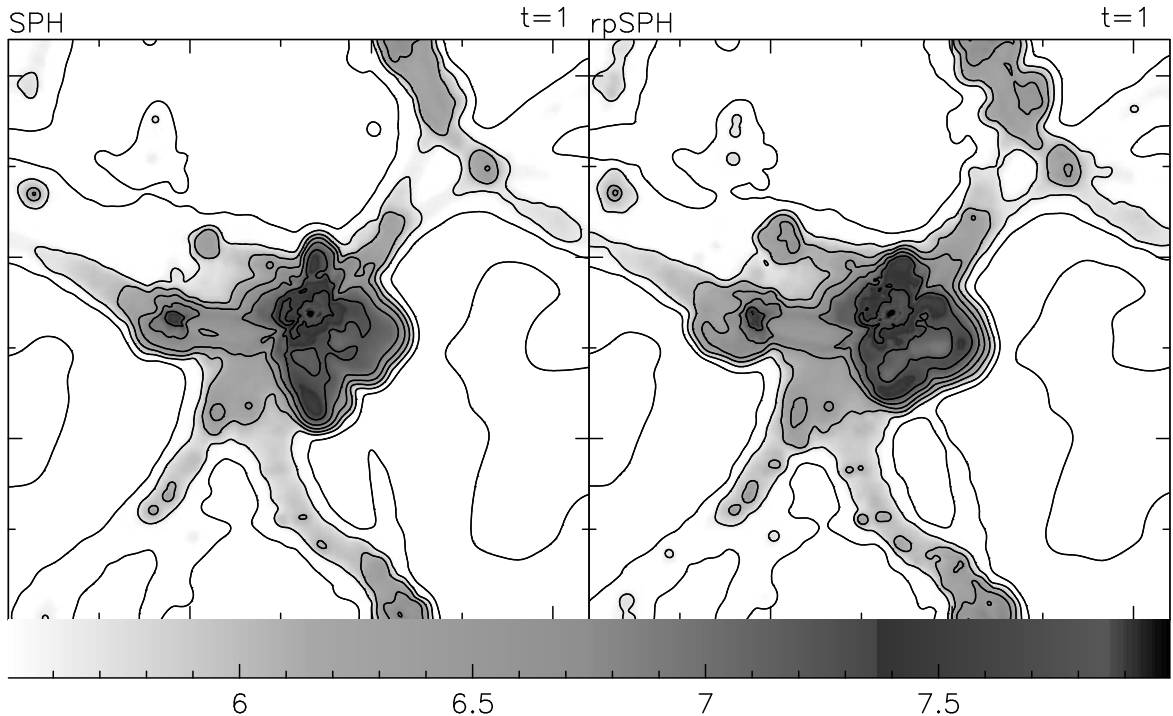}
\vspace{-12.cm}
\caption{Logarithm of temperature in a two dimensional slice at
  redshift zero derived form the initial conditions of the Santa
  Barbara Galaxy Cluster comparison project
  \citep{1999ApJ...525..554F}. The plots are 64 Mpc across and show 10
  contours within the interval from 30,000 to one hundred million
  degrees Kelvin.  The SPH and {\em rpSPH} simulations used $128^3$
  particles and 300 neighbors with a tolerance of one, $\alpha=3$. The
  agreement is excellent. Subtle differences in the postshock
  temperature can be seen. {\em rpSPH} seems clearly robust enough for
  large scale cosmological integrations.
}\label{fig:SB-temps}
\end{figure*}

\begin{figure}
\includegraphics[width=0.49\textwidth]{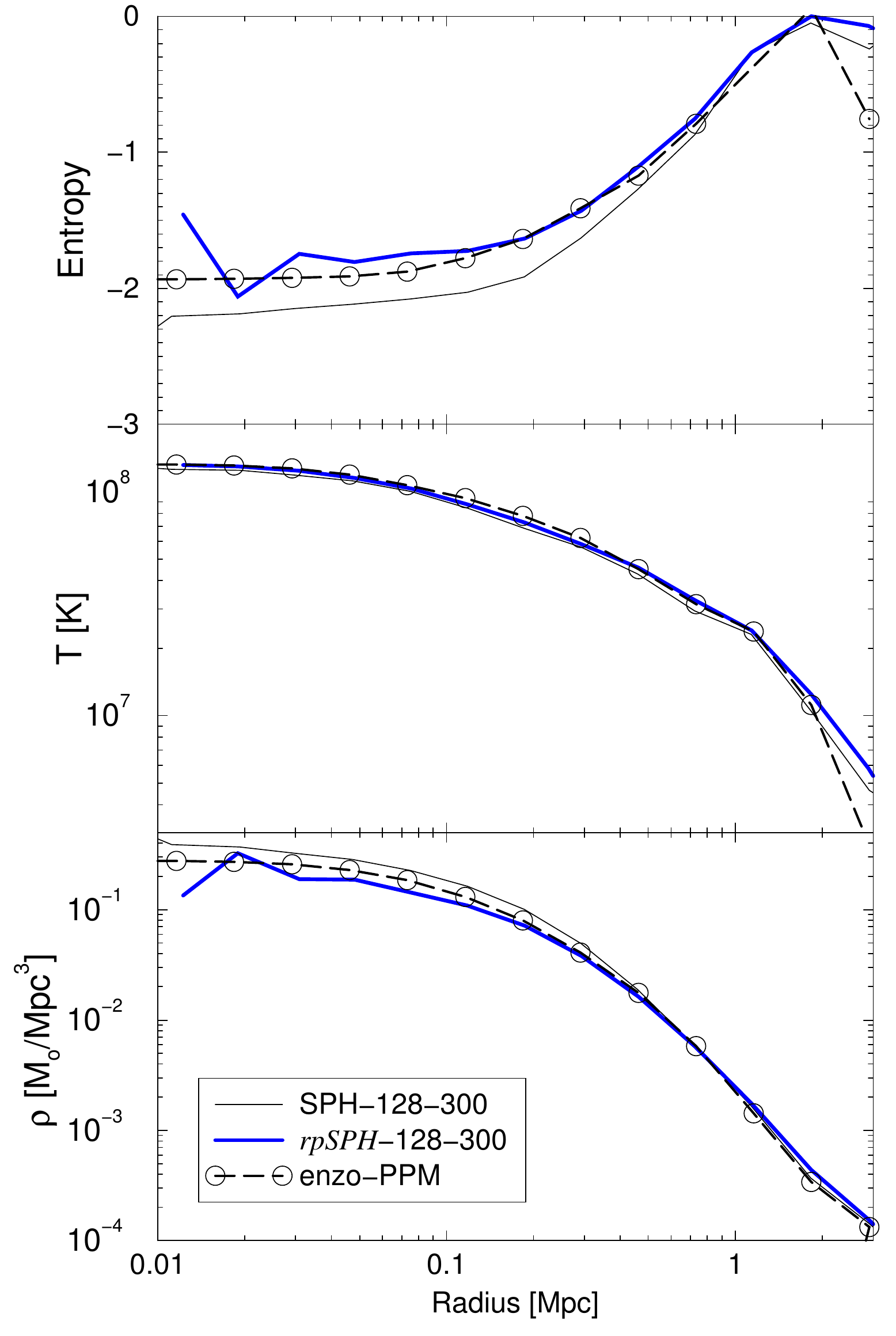}
\caption{Spherically averaged profiles of density, temperature and
  entropy of the Santa Barbara Galaxy Cluster comparison project
  \citep{1999ApJ...525..554F} The SPH and {\em rpSPH} simulations used
  $128^3$ particles and 300 neighbors with a tolerance of one,
  $\alpha=3$, and are compared to results from an AMR calculation
  using the piecewise parabolic method (enzo-PPM). The agreement is
  excellent. Its interesting that {\em rpSPH} gives higher central
  entropy as standard SPH more comparable to the grid code as this has
  been the subject of much discussion in recent years. Note also that
  the entropy profiles of {\em rpSPH} and the AMR calculation agree
  better well away from the central resolution limited part at radii
  between 0.1 and 1. 
}\label{fig:SB-comp}
\end{figure}

In 1995 a comparison project was initiated that aimed to compare all
numerical cosmology codes at that time for relevant realistic initial
conditions. The study focused on three dimensional calculations of the
formation of a galaxy cluster in the standard CDM scenario of
structure formation. The choice was a setup which does not include any
other physics than cosmological hydrodynamics with an ideal gas
equation of state (often referred to as adiabatic simulations despite
the entropy generation in shocks). The study produced a detailed
report in \cite[][F99, herafter]{1999ApJ...525..554F}. One of the most
surprising findings of the study was that while there was very good
agreement between the six different SPH codes used in the study they
did not agree with the solutions of the grid based codes. The central
entropy of the simulated galaxy cluster differed markedly between the
grid and SPH codes. In particular, the only AMR code in the study by
\citet{1997ASPC..123..363B} which is now called {\em Enzo}
\citep{Bryan:2001,2004astro.ph..3044O} found a flat entropy core while
the particle codes found the central entropy to continue to rise
towards smaller radii. Note that there has been a significant debate on what the
correct solution may be and potential sources between the differences
between grid and particle based methods \citep{2005MNRAS.364.1105S,
  2005MNRAS.364..753D, 2009arXiv0902.4002K, 2008MNRAS.387..427W,
  2009MNRAS.395..180M, 2007MNRAS.380..963A, 2010MNRAS.401..791S} is
the real reason for the difference.

We do not attempt a resolution study here but simply show how an {\em
  rpSPH} solution compares to an SPH run with otherwise identical
parameters and the solution derived with a cosmological AMR code. We
use $128^3$ gas as well as dark matter particles for the particles
based approach. For the AMR code we again use {\em Enzo} already used
in F99 using $128^3$ dark matter particles a root grid of $128^3$
cells and seven additional refinement levels. Refinement is based on
density thresholds in the baryons and dark matter component.  The
viscosity parameter in the particle based runs was $\alpha=3$, and a
neighbor number of 300 with tolerance of one was used. The initial
redshift is $z=63$ and the initial temperature $10^9$ Kelvin.

A two dimensional slice through the temperature field is given in
Figure~\ref{fig:SB-temps} comparing SPH to {\em rpSPH}. Only
relatively subtle differences are found. There are perhaps slightly
more small scale features visible in the {\em rpSPH} calculation and
some slight differences in the post shock gas in the main cluster are
visible. Slightly more shocking occurs at larger radii towards low
density voids in the {\em rpSPH} vs. SPH calculation.

Figure~\ref{fig:SB-comp} compares the solutions using spherically
averaged radial profiles as described in F99. The entropy profiles of
{\em rpSPH} agree better with the AMR than the SPH results. The {\em
  rpSPH} solutions shows the lowest central densities of the three
methods and agrees better with the slightly shallower density profile
of the grid code. Clearly the differences between all three, however,
are rather subtle given that one evolved this system for 13 billion
years which are many tens of sound crossing times for the central part
of the resulting cluster.

Both SPH and {\em rpSPH} simulations have a final linear momentum
corresponding to 4.1 and 3.2 km per second per gas particle,
respectively. The difference vector between the final gas momenta of
the simulation has a magnitude of 2.6 km/s per gas particle. This is a
difference of order one half of a percent of the mass weighted mean
r.m.s. velocity of $\sim 400$ km/s. Obviously, giving up the strict
linear momentum conservation in our equation of motion has not lead to
any noticeable difference in this measure but has improved the
comparison with results from adaptive mesh refinement codes. 

However, this particular application is relatively easy as dark matter
dominates the gravitational potential. As we will show further below
{\em rpSPH} is quite easy to break with self-gravitating fluids.

\section{Further Benefits of {\em rpSPH}}

\subsection{Variable masses}\label{sec:mm}

\begin{figure}
\includegraphics[width=0.5\textwidth]{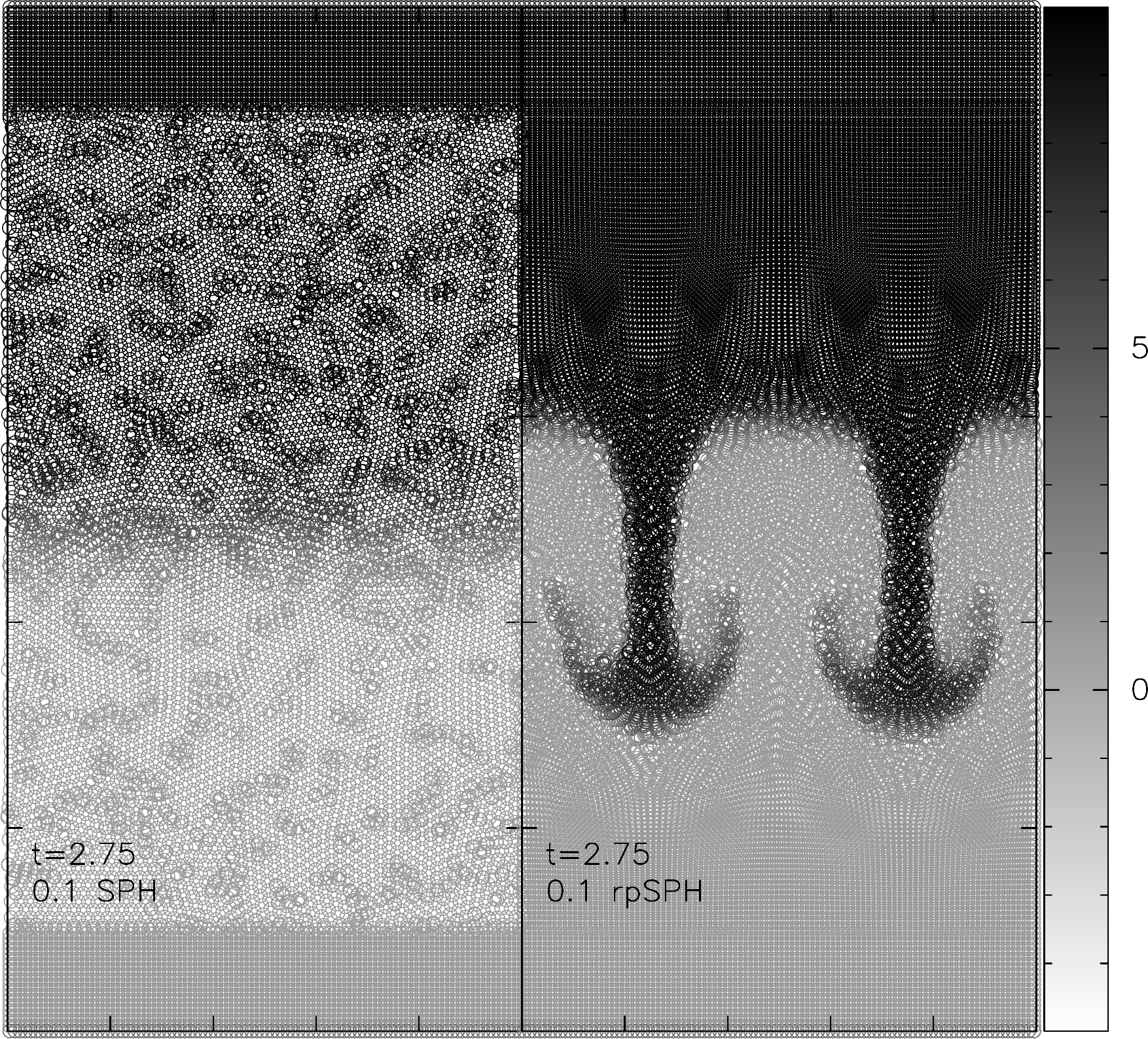}
\caption{Density in a Rayleigh--Taylor test with a ten times heavier
  fluid on top than the bottom. Both SPH and {\em rpSPH} runs used an
  initial uniform grid of 100 by 200 particles with varying masses to
  describe the higher density for the interface and top fluid. The
  clumping instability of standard SPH creates the pattern on the left
  while {\em rpSPH} works perfectly fine. Being able to do accurate
  calculations even with largely disparate particle masses at low
  resolution will be an enormous benefit in many situations where
  large density contrasts are observed but one needs to retain high
  accuracy in the low density regions. }\label{fig:mm-RT}
\end{figure}

Next we demonstrate that using our pressure force discretisation give
another very important advantage. Simulations with drastically varying
particle masses continue to give correct results. This is markedly
different compared to previous SPH simulations employing particle
splitting. The latter only worked reasonably as long as different
particle masses were very well separated spatially. As an explicit
example we revisit the Rayleigh--Taylor problem from above. This time
we initialize particles on a uniform lattice and model the density
contrast by changing the particle masses according to the density
profile. We employ a density at the top ten times the one of the one
at the bottom fluid to demonstrate that this is not just a marginally
better aspect of {\em rpSPH}. Figure~\ref{fig:mm-RT} compares the
SPH and the rpSPH solution again for $\delta v=0.01$. Instead of
showing the density field we show the particles painted by circles and
colored by their density. This gives us an opportunity to see that
{\em rpSPH} does not show any clumping instability while it is severe
for SPH. In this run we only use 100 by 200 particles demonstrating
that {\em rpSPH} handles the Rayleigh--Taylor problem very well at
small initial perturbations and low particle numbers.

The Sedov-Taylor blast wave above was also carried out with a staggered
grid of particles of varying mass and retains a nice spherical shape
despite the multiple squares introduced in the staggered ``mesh'' of
the initial particle distribution.

It is of great interest for a method to be stable under largely
varying particle masses. If it is one can use particle splitting
likely without worrying too much of how to place the new particles and
keep them separate from particles with different masses
\citep[e.g.][]{2002MNRAS.330..129K}.

\subsection{Formulation for Magnetic Forces}

There have been many attempts to implement ideal MHD into the standard
SPH formalism \citep[e.g.][]{1977MNRAS.181..375G, 1985MNRAS.216..883P,
  2004MNRAS.348..123P, 2009MNRAS.398.1678D} In our experience {\em
  rpSPH} performs in the hydro part better than the Morris
formulation. The latter has been used in implementing ideal MHD into
SPH \citep{Morris1996, 2004MNRAS.348..123P, 2009MNRAS.398.1678D}. 
Therefore, we expect that our new discretization may be of use in this
case as well. 

A symmetric conservative form of the Lorenz force is generally
implemented using the magnetic stress tensor \citep{1985MNRAS.216..883P},
\begin{equation}
M_i^{kl} = \left( \vec{B}_i^k\vec{B}_i^l - \frac{1}{2}|\vec{B}_i|^2\delta^{kl}\right).
\label{eq:maxtens}
\end{equation}
So the acceleration from the magnetic fields on the $i$-th particle
is then written as
\begin{eqnarray}
\left(\frac{\mathrm{ d}\vec{v}_i}{\mathrm{ d}t}\right)^{(\mathrm{B})} = \frac{1}{\mu_0}
\sum_{j=1}^{N}m_j \left[f_i\frac{M_i}{\rho_i^2}
                       \cdot\vec{\nabla}_i W_i \right.   \nonumber
+
                       \left.f_j\frac{M_j}{\rho_j^2}
                       \cdot\vec{\nabla}_j W_j \right].
\end{eqnarray}
In the limit where we only have forces from the magnetic pressure
gradient only the diagonal of the tensor has terms which are $B_i^2/2$
and we recognize this discretisation as exactly the form of
equation~(\ref{eqnmot}) above. So also the Lorentz force lends itself
to be discretised following our new approach. We split the force into
the tension component and the magnetic pressure component without loss
of generality into
\be \rho {\partial v \over \partial t} =  -\nabla {\vec{B}^2 \over 8\pi} + {(\vec{B}\cdot\nabla)\vec{B}
  \over 4\pi}.
\ee
In both terms one takes spatial derivatives and hence is allowed to
subtract a constant. Choosing $\vec{B_i}\cdot\vec{B_i}$ for the first
one and $\vec{B_i}$ for the second one avoids finite pairwise forces
between particles in regions of constant field. So the isotropic part 
becomes 
\be
\frac{\mathrm{ d}\vec{v}_i}{\mathrm{ d}t} = -\frac{1}{8\pi}
\sum_{j=1}^N m_j   f_j\frac{\vec{B}_j^2-\vec{B}_i^2}{\rho_j^2}
                       \vec{\nabla} \bar{W},
\ee
to which the tension force is added
\be
\ \ \ \ \ + \frac{1}{\mu_0}
\sum_{j=1}^{N}m_j  f_j\frac{\vec{B_i}\cdot (\vec{B}_j-\vec{B}_i)}{\rho_j^2}
                       \vec{\nabla} \bar{W}.
\ee

Both the terms simply add to the accelerations in the force
calculation. All that is left to do is to replace the fastest signal
velocity with the one given in equation~(46) of
\cite{2004MNRAS.348..123P} and one has a MHD implementation of SPH.
On some simple initial tests this formulation seems to work quite well
even without any regularization technique \citep[e.g.][and references
therein]{2009MNRAS.398.1678D} or artificial B field dissipation. A
full exploration of the performance of this discretization though is
beyond the scope of this contribution.

\section{How to break {\em rpSPH}}

From the tests above it is clear that at least in the weakly
compressible limit {\em rpSPH} is a very useful improvement over the
standard formulation. However, giving up momentum and energy
conservation is a big price to pay for those advances and {\em rpSPH}
cannot possibly replace standard SPH in all problems of interest. {\em
  rpSPH} should be easy to break at low resolutions when one does not
resolve the pressure gradients adequately. We now give an illustrative
example that makes {\em rpSPH} give very bad results which shall serve
as a cautionary note and things to look for when applying {\em rpSPH}.

The Evrard collapse of a cold gas sphere \citep{1988MNRAS.235..911E}
has been extensively used for verification of astrophysical SPH codes
\citep[e.g.][]{2004NewA....9..137W, 2005MNRAS.364.1105S}. In the
version that is part of the Gadget distribution it is realized with
only 1472 equal mass particles. A centrally concentrated cloud of cold
gas collapses, bounces and eventually virializes. Vacuum boundaries
are assumed.  Very clearly this is only meant to investigate energy
conservation of the code and is a simple test running in seconds on
ones laptop.

\begin{figure}
\includegraphics[width=0.47\textwidth]{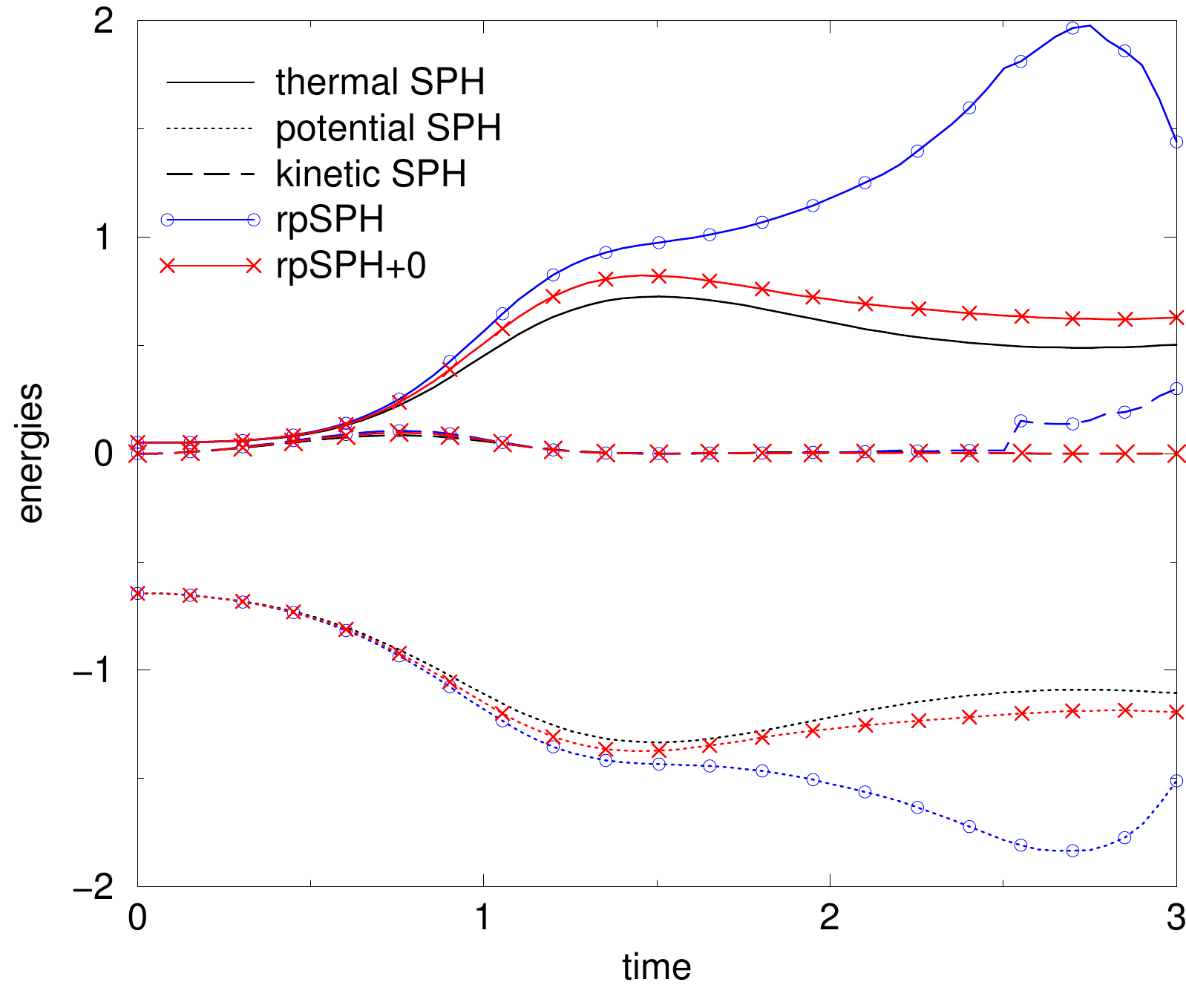}
\caption{Energies in the Evrard gas collapse test. }\label{fig:Evrard-energies}
\end{figure}

Figure~\ref{fig:Evrard-energies} shows the standard SPH solution
together with a completely failed solution of {\em rpSPH}.
We should not that as we increase the resolution {\em rpSPH} does
converge to the same solution as standard SPH. However, this is a
clear example where too little resolution coupled with a solver that
does not conserve momentum or energy will fail completely. 

Figure~\ref{fig:Evrard-energies} also gives the energies in the Evrard
collapse for which we formally added a zero and label it {\em
  rpSPH+0}.  The term added to the momentum equations right hand side
is $P/\rho\, \nabla 1 = 0$, which in SPH reads \citep[e.g.][]{2001MNRAS.323..743R}
$$
  \sum_{j=1}^N m_j \left[
  \frac{P_i}{\rho_i \rho_j } {\nabla_i W_{ij}(h_i)} \right]. 
$$
This term which tends to zero when the density and pressure gradients
are well resolved. For our poor resolution setup, however, we see that
it gives better energy conservation. We do not recommend this
formalism over {\em rpSPH}, however, since it still has the problems
with contact discontinuities and weakly compressible flows. However,
it highlights that as for any numerical study resolution studies
are crucial and that {\em rpSPH} will likely be of little use when one
cannot afford sufficient resolution for the particular problem one is
interested in.

\section{Conclusions}

We have presented a novel discretization of the pressure equation for
the smoothed particle hydrodynamics, which we call {\em rpSPH} that
removes the local pressure from the scheme and only considers pressure
gradients. This methodology avoids the clumping and banding
instability, artificial surface tension, unphysical particle noise,
dramatically reduces inherent shear viscosity and numerical
dissipation, and allows to realistically evolve density distributions
sampled even with disparate particle masses. We have discussed a large
number of test in all of which our new discretization outperforms the
traditional SPH results. While our approach is not manifestly momentum
conserving and easy to break wiht self-gravity and or low resolutions
it clearly seems much more accurate than previous approaches to
Lagrangian hydrodynamics using SPH. Since our formulation is more
accurate and requires as little as one line of code to be changed in
previous implementations we do believe it to likely to be useful. 
The caveat of {\em rpSPH} remains that if
one knows that one cannot afford to resolve the pressure gradient in
ones initial conditions that because it is not momentum conserving it
can give very wrong results. Fortunately, a resolution and convergence
study can reveal whether one is in this limit.

In summary, some of the biggest shortcomings of SPH can in some
circumstances can be overcome if
one gives up the idea of applying equal but opposite forces to
particle pairs. While the latter is what happens physically to the
atoms or molecules making up the fluid it is simply incorrect for
Lagrangian fluid elements the particles are meant to represent.
Physically it also does not make sense to introduce repulsive forces
for two spatially separated points even when there is no pressure
gradient between them. To require such symmetry between particles
neglects that they are spatially separated and that the gradient of
the pressure field is different at the two locations in general.  Our
new discretization avoids these unphysical forces and allows the SPH
particles to behave as Lagrangian volume elements recovering fluid
behaviour in a large number of tests. 

We have successfully used a fifth order spline kernel giving smaller
errors on the uniform shear problem and the Rayleigh-Taylor problems
discussed above. Consequently, we believe that further improvements to
{\em rpPSPH} should be possible in the future. 

We have also studied multiple forms of discretising the specific
internal energy equation \be \frac{d\epsilon}{dt} = - \frac{P}{\rho}
  \nabla \cdot \vec{v} \label{equ:ie}.\ee The simplest version that
we successfully applied to some of our test problems is given by
\be \frac{d\epsilon}{dt} \approx \sum_{j=1}^N  
\frac{m_j}{\rho_j}  \frac{P_j}{\rho_j} \frac{(\vec{x}_i - \vec{x}_j) \cdot
  (\vec{v}_i-\vec{v}_j)}{|\vec{x}_i - \vec{x}_j|} {\nabla_i W_{ij}(h_i)}.
\ee
While we prefer the entropy formulation this form here may be useful
for codes that start from an internal energy formulation. 

While standard SPH is conservative it fails to correctly capture fluid
instabilities and shows large non-Newtonian viscosity.  {\em rpSPH},
on the other hand is more accurate, but is not inherently momentum
or energy conserving. Consequently it is a useful modification to the
SPH algorithm when one is studying problems where one can afford to
resolve the relevant pressure gradients and the density field.

\section*{Acknowledgements}
I would like to acknowledge the hospitality of the Zentrum f\"ur
Astronomie Heidelberg and the Institut f\"ur Theoretische Astrophysik
in particular where this work was started. I am particularly indebted
to Volker Springel for making Gadget public making this work possible
in the first place, and thank Paul Clarke, James Wadsley and Ralf
Klessen and my group and colleagues at KIPAC for useful discussions.
Furthermore, I am indebted to Landesstiftung Baden W\"urttemberg for
travel support.  This work was partially supported by NASA ATFP grant
NNX08AH26G and NSF AST-0807312.  This research has made use of NASA's
Astrophysics Data System Bibliographic Services and {\em splash}
developed and described by \citet{2007PASA...24..159P}.

\appendix
\section{Shearing flows of Uniform Density with standard SPH}\label{sec:uni}

While {\em rpSPH} overcomes the problems found for shearing flows this
appendix is more of historical interest. However, some readers may
find it worthwhile since to our knowledge the surprisingly large shear
viscosity and its erratic behaviour with increasing particle numbers
has not been documented previously. 

A simple two dimensional setup uses an adiabatic index of $\gamma=1.4$
in a unit square domain $x\in \{0,1\}$, $y\in \{0,1\}$ with periodic
boundary conditions. All particles are set up on an exactly square
lattice with a density of $\rho(x,y)=1$ so that the initial density
estimate from the SPH kernel in fact gives a density estimate of unity
to better than four parts in one thousand. We then add different
velocity perturbations to this uniform distribution. We set the
pressure to $P_0=\rho/\gamma$ to have a sound speed of unity. For the
first tests here we only use $50^2$ particles as there are no features
to resolve. In all cases we evolve to time $t=4$.

First we start with no velocity perturbation. I.e. a completely static
uniform density distribution evolved over four sound crossing
times. Using 30 neighbors the density estimate by all particles is
$1.00345$. For different neighbor numbers this fluctuates around 1 and
is close enough. We will use 30 neighbors for most of the rest of this
section.  Initially all velocities are zero yet after four crossing
times we have an r.m.s. velocity $v_{rms}=\sqrt{\frac{1}{N}\sum_N
  v_x^2+v_y^2}\approx0.01$ with no obvious preferred direction. This
results is obtained for the typical viscosity value of $\alpha=1$. For
lower values this random noise increases to $v_{rms}\approx 0.036$ for
$\alpha=1/10$. So clearly even under the most quiet conditions
imaginable, a uniform density in pressure equilibrium, we could not
represent velocities of order a few percent of the sound speed. Now
let us perturb the velocity along the $x$ direction and set it to a
uniform value of $1$. This should be exactly identical to the previous
setup given that SPH is formulated to be Galilean invariant. Now the
random noise $v_{rms}=\sqrt{{N_p^{-1}}\sum_{N_p}
  (v_x-1)^2+v_y^2}\approx0.009$ for $\alpha=1$ and again $v_{rms}
\approx 0.035$ for $\alpha=1/10$. 

Now for our next experiment with this uniform density setup we use
$v_x(y) = \delta v_{\rm y}\cos(2\pi\, y)$ with $\delta v_{\rm
  y}=1/2$. This shear flow setup gives an average kinetic energy of
$1/8$. After only 4 sound crossing times (or two crossing times of the
fastest particles) the mean kinetic energy of particles has decreased
by 15\% and the r.m.s. velocity in $y$ direction is already $\approx
0.063$ when using a viscosity parameter of $\alpha=1/10$. Using the
standard value $\alpha=1$ we have a lower r.m.s. velocity in the $y$
direction of $\approx 0.032$ yet at the same time the total kinetic
energy has decreased by a as much as 27 percent suggesting that the
standard value does convert unacceptable levels of the shear into
heat. This hardly is inviscid flow! Again for $\alpha=1$ but $200^2$
particles which allow the shear to be better resolved one would hope
for less dissipation. Yet we find that the total kinetic energy still
decreases by 30\% and the r.m.s. velocity in the y direction becomes
$\approx 0.021$. We have also run this test with $300^2$ particles and
find that the kinetic energy dissipation is approximately independent
of resolution up to this particle number.  The kinetic energy lost
after two crossing times was 30.3\% and the final r.m.s. velocity
fluctuations in the y-direction was $\approx 0.019$ i.e. a fiftieth of
the sounds speed. The latter velocity dispersion became as high as a
thirtieth during the first time interval $t\sim 0.2$, decreased and
then stayed stable afterwards at $\approx 0.02$. The maximal vertical
velocities are as much as one tenth of the sound speed for a problem
which should not develop any perpendicular velocities.

Perhaps using a lower viscosity parameter could help? So with
$\alpha=0.1$ and $200^2$ particles this indeed gives lower overall
dissipation of the kinetic energy of ``only'' 16\% but then gives
random motions perpendicular to the shear of $\approx 0.030$.
Figure~\ref{fig:reduced-S} summarizes the loss of kinetic energies for
simulations with $50^2$ and $100^2$ particles for 30 and 48 neighbors
and the artificial viscosity parameters of $\alpha=0.1$ and
$\alpha=1$.

From the figure it is very clear that using more particles actually
leads to {\em more} dissipation. How is one supposed to carry out a
resolution study when the numerical dissipation can keep increasing
when using more and more computational resources?

A histogram over all particles showing their $x$ velocity as a
function of their $y$ coordinate in Figure~3 reveals how strongly the
shear viscosity turned the initial sinusoidal perturbation into
flattened extrema with linear profiles between them. This graph looks
similar for different viscosity values. The bottom panel of Figure~3
visualizes the particles making the ones that received the entropy
clearly visible. The lowest (initial) value one should expect is white
and in fact below the minimum on that image
$P/\rho^{\gamma}=1/\gamma\approx 0.7143$. However, one can see clear
bands in the places where one finds the largest gradients in the shear
velocities. The clumping instability is clearly visible through the
bunching of entropy values in the plot.

It is worth noting that the Balsara switch \citep{1995JCoPh.121..357B}
which is designed to limit this shear dissipation indeed helps.
Without it we find that after two crossing times 48\% of the kinetic
energy are already artificially dissipated in the $200^2$ test with
the sinusoidal shear at Mach one half and a uniform density. These
48\% are to be compared to the 30\% which were dissipated using the
Balsara switch. 

\begin{figure}
\includegraphics[height=0.44\textwidth]{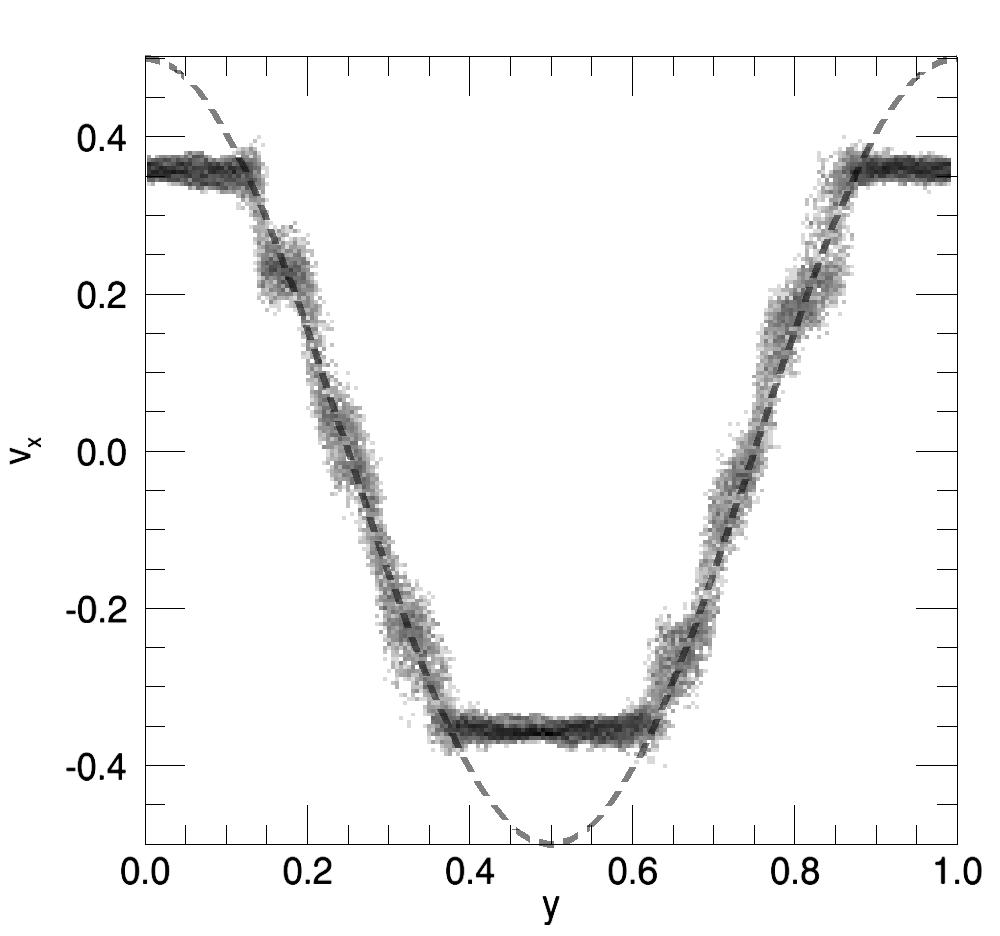}\hspace{\myfix}
\includegraphics[height=0.44\textwidth]{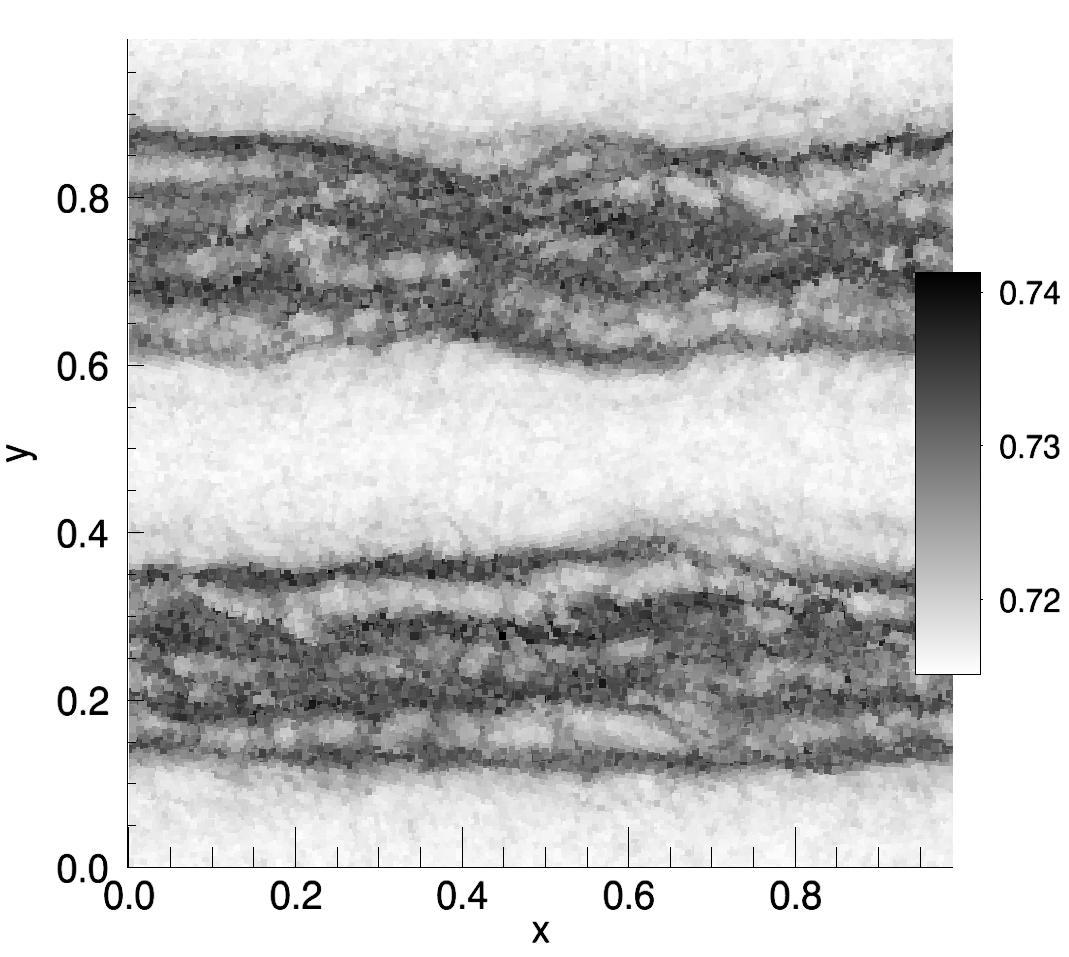}\hspace{\myfix}
\caption{Uniform density shear test. {\bf Top:} The solid line gives
  the amplitude of the uniform initial $x$ velocity modulated as a
  function of $y$. The underlying histogram is the particle
  distribution after four sound crossing times when using a viscosity
  parameter $\alpha=1$. It looks slighlty better for the lower
  artificial viscosity parameter of $\alpha=1/10$ but leads to larger
  vertical velocity perturbations. {\bf Bottom:} Entropy of the
  particles in a $200^2$ run showing bands of the material that
  received the entropy ($P/\rho^\gamma$) in the 200$^2$ run with
  $\alpha=1$. The values are plotted as squares with side lengths of
  half the SPH smoothing length. }\label{fig:uni}
\end{figure}

The viscosity limiter implemented in Gadget-2 does not influence the
results here. Also decreasing the Courant factor by an order of
magnitude can change the exact amount of dissipation but does in
general not decrease it appreciably.

That the effective shear viscosity changed little with increasing
particle number is very unfortunate. However, if we increase the
neighbor number employed to 60 neighbors the effective shear viscosity
drops dramatically and only 1\% of the total kinetic energy is
artificially dissipated over the same time interval. However this
comes at the price of particles clumping into bands through the well
known tensile instability.  As discussed in some detail by
\citep{2009arXiv0906.0774R} the amount of clumping is specific to the
kernel choice.

One compromise for this uniform shear problem is a neighbor number of
48 which leads to banding and only about 3\% kinetic energy
dissipation in the two crossing times. Simple scaling implies then a
choice of $48^{3/2}\approx 333$ neighbors for three dimensional
calculations. A number much larger than typically employed.

We have evolved this same test to many more crossing times and find
that at the larger neighbor numbers dissipation simply occurs later
but in fact looks qualitatively just like in the low neighbor number
case. This enables now a further discussion of the origin of this
artificial shear viscosity. In the top panel of Figure~\ref{fig:uni}
we can see how the extrema in the velocity are clipped by the
viscosity and that the particles positions which were one a regular
lattice in the $y$ direction spread out and led to banding. This in
large part comes from the non-zero $y$ velocity the particles obtain
leading them to artificially mix into regions perpendicular to the
velocities they have. So small fluctuations in the pressure forces
enable a coupling taking energy from the $x$ velocity to stimulate
motions in the $y$ direction. Particles then artificially mix into
regions of the flow where they start to interact with fluid parcels of
different shear velocities triggering the artificial viscosity which
then tries to damp this noise. This is why smaller artificial
viscosity parameters will lead to larger r.m.s. $y$ velocities. This
also explains why increasing the neighbor number delays this
artificial shear viscosity in that it decreases the amplitude of the
forces leading to the $y$ velocities. 

Interestingly the uniform density tests presented here are similar as
the one presented by \cite{2006MNRAS.365..199M} in terms of the
velocity profile and in the sense that it is a low mach number
flow. \cite{2006MNRAS.365..199M}, however, chose to pick $\gamma=7$
making the EOS very stiff. While this may be a useful trick to model
incompressible flow with SPH it is not something we repeat here since
for most applications in astrophysics we have $1\simlt \gamma \simlt
5/3$. However, even with the stiff equation of state his Figure~2
shows the same clipping of the maximal velocity amplitudes as our
Figure~\ref{fig:uni} after only one crossing time. This agrees with
our findings that only for short time scales (as compared to the
crossing time) the shear viscosity may be negligible. We just differ
in the interpretation of whether this is an acceptable level of
dissipation or not. 

% A seemingly similar test was presented by \cite{2002ApJ...569..501I}
% but in the highly supersonic limit. This is also a very important
% regime, in particular in the context of thin accretion disks in star,
% planet and black hole formation. However, discretization noise can
% quickly lead to very large density variations exciting a large range
% of modes giving a nonlinear solution far from ones initial homogeneous
% expectation. It is a pity that none of these studies report the number
% of neighbors they employed. We learned above that this parameter is
% even more relevant than the viscosity parameter chosen. Perhaps the
% large discrepancies in those two studies on the same test could be
% explained if \cite{2002ApJ...569..501I} had used significantly fewer
% neighbors than \cite{2006MNRAS.365..199M} and/or perhaps let the
% number of neighbors vary dramatically during their calculations.

\section{Modifying Gadget-2.0.4 to rpSPH}

For the convenience of other researchers we give the details of what to do to
convert Gadget-2.0.4\footnote{\tt http://www.mpa-garching.mpg.de/gadget/}
 to take advantage of the {\em rpSPH}
discretization. In \verb hydra.c  find 
the line that reads
% {\small
% \begin{lstlisting}
% soundspeed_j = sqrt(GAMMA * p_over_rho2_j 
%                     * SphP[j].Density);
% \end{lstlisting}
% }
% and replace it with 
% {\small
% \begin{lstlisting}
% soundspeed_j = sqrt(GAMMA * SphP[j].Pressure 
%                     / SphP[j].Density);
% \end{lstlisting}}
% then find 
% {\small \begin{lstlisting}
% p_over_rho2_j = SphP[j].Pressure  
%      / (SphP[j].Density * SphP[j].Density);
% \end{lstlisting}}
% and change it to
% {\small \begin{lstlisting}
% p_over_rho2_j = (SphP[j].Pressure - pressure)
%     / (SphP[j].Density * SphP[j].Density);.
% \end{lstlisting}}
% Now we only have to change 
{\small \begin{lstlisting}
hfc = hfc_visc + P[j].Mass*(p_over_rho2_i*dwk_i 
       + p_over_rho2_j*dwk_j)/r;
\end{lstlisting}}
and change it to 
{\small \begin{lstlisting}
hfc = hfc_visc+P[j].Mass/SphP[j].Density*
 (SphP[j].Pressure-pressure)/SphP[j].Density*
 dwk_i/r;
\end{lstlisting}}
and the conversion is complete. 

Another form that fits more closely to the artificial viscosity
prescription useful for problems with large density gradients is
{\small \begin{lstlisting}
hfc = hfc_visc+P[j].Mass/SphP[j].Density*
 (SphP[j].Pressure-pressure)/SphP[j].Density*
 (dwk_i+dwk_j)/2/r;
\end{lstlisting}}

In order to keep standard SPH for very strong shocks matching the
standard viscosity implementation one may choose to keep both lines
but preface the latter with 
{\small \begin{lstlisting}
    if  (-h_i*divVel < 3.*soundspeed_i) 
\end{lstlisting}}
, or other criteria that trigger at strong shocks.

\bsp
\bibliography{Old-MyReferences}{}
\bibliographystyle{astron} 

\label{lastpage}

\end{document}